# A geometric framework for spin relaxation

Sophia N. Fricke [a*], Kieren A. Harkins [b], Ashok Ajoy [b,c], Jeffrey A. Reimer [d]

[a] Pines Magnetic Resonance Center
University of California, Berkeley
Berkeley, CA 94720, USA

[b] Department of Chemistry
University of California, Berkeley
Berkeley, CA 94720, USA

[c] Chemical Sciences Division
Lawrence Berkeley National Laboratory
Berkeley, California 94720, USA

[d] Department of Chemical and Biomolecular Engineering
University of California, Berkeley
Berkeley, CA 94720, USA

[*] Corresponding Author, **Email:** snfricke@berkeley.edu

**Author contributions:** S.N.F. designed, performed, and analyzed research and wrote the paper, K.A.H. performed research and edited the paper, A.A. edited the paper, and J.A.R. directed the research and edited the paper.

**Abstract**

Spin relaxation is conventionally described by two independent phenomenological rates – longitudinal ($R_1$) and transverse ($R_2$) – whose separation obscures a deeper structural unity. Here we develop a geometric framework in which dissipation is represented by a single covariant relaxation tensor acting in Liouville space, from which $R_1$ and $R_2$ emerge as complementary projections. This tensor structure is not merely formal but is experimentally accessible through pulse sequences that probe noncommuting directions in spin space. Using hyperpolarized $^{13}C$ spins in diamond with nitrogen-vacancy centers, we show that commuting pulse trains yield effective relaxation matrices that are approximately diagonal, while noncommuting sequences produce off-diagonal components that vary with transmitter frequency offset and pulse ordering, providing evidence that relaxation is a directional process governed by a tensor rather than a pair of scalar rates. Complementary measurements of geometric phase demonstrate that noncommuting dynamics introduce ordering-dependent effects that are separable from dissipation, consistent with the interpretation of relaxation as geometric transport on the state manifold. This framework unifies Bloch, Redfield, and Lindblad descriptions within a coordinate-independent formulation and provides a natural language for relaxation in driven, anisotropic, and non-equilibrium spin systems.

## Introduction

Spin relaxation governs the return of a perturbed spin system to equilibrium and underlies the contrast mechanisms, sensitivity limits, and coherence lifetimes that define magnetic resonance experiments. Its conventional description separates two phenomenological processes: longitudinal relaxation ($R_1$), associated with energy exchange between spins and the lattice, and transverse relaxation ($R_2$), associated with the dephasing of spin coherences. This separation has proven enormously productive, organizing the interpretation of relaxation data across spectroscopy, imaging, and materials characterization, as well as providing the foundation for connecting measured rates to microscopic dynamics through spectral density functions.[1–3]

Yet the separation is an approximation, and its limitations become apparent in regimes that are increasingly central to modern magnetic resonance. For example, in driven systems, where time-dependent control fields create effective Hamiltonians that vary with the pulse sequence, relaxation rates inferred from different protocols yield protocol-dependent values that a scalar $R_1/R_2$ description cannot reconcile.[4–10] In anisotropic or spatially heterogeneous materials, the assumption that longitudinal and transverse processes are decoupled along fixed laboratory frame axes may not hold. And in hyperpolarized systems, where initial states are far from thermal equilibrium, and in strongly driven systems where the secular approximation breaks down, the scalar description leaves the relationship between coherence loss and energy exchange implicit rather than manifest.[11–14]

The shared microscopic origin of $R_1$ and $R_2$ is already implicit in Redfield theory, where both rates arise from the same spectral densities of environmental fluctuations (an explicit description is provided in SI §2.8), and their interdependence is encoded in the relaxation superoperator. More generally, Gorini–Kossakowski–Sudarshan–Lindblad (GKSL) and Lindblad formalisms describe open quantum system dynamics in terms of completely positive maps on the density operator, providing a framework within which both rates emerge from a single dissipative structure.[15–20] Yet these approaches are typically expressed in specific representations and emphasize different aspects of the dynamics. Redfield theory is formulated in a fixed eigenbasis of the static Hamiltonian and emphasizes microscopic mechanisms; GKSL theory guarantees complete positivity but is usually written in a particular operator basis. As a result, the shared structure linking longitudinal and transverse processes is often left obscured. What has been lacking in the magnetic resonance context is a formulation that makes this unity explicit, coordinate-independent, and experimentally accessible – one that recovers the standard Bloch picture as a special case while accommodating the richer dynamics that arise under driven or anisotropic conditions, and connecting to the geometry of open-system dynamics that has been developed more abstractly elsewhere.[20–22]

Here we develop a geometric framework for spin relaxation in which dissipation is represented by a covariant relaxation tensor acting in Liouville space. $R_1$ and $R_2$ emerge as complementary projections of this tensor onto a chosen observable basis; the tensor is the fundamental object, and what any experiment accesses is a control-dependent projection of it. The framework joins Bloch, Redfield, and Lindblad descriptions within a common coordinate-independent formulation and provides a language for describing relaxation as geometric transport on the state manifold. Crucially, the tensor structure is not merely a theoretical convenience: it carries observable consequences. Pulse sequences that traverse noncommuting directions in spin space probe different projections of the underlying tensor, producing trajectory distortions, off-diagonal relaxation matrix components, and ordering-dependent dynamics that cannot be accounted for by scalar rates. We demonstrate these effects experimentally using $^{13}$C nuclear spins in diamond, hyperpolarized via optical pumping of nitrogen-vacancy (NV) centers, and show that geometric phase, which is generated by the coherent (control) part of the Liouvillian, accumulates independently of dissipation under cyclic noncommuting sequences, providing a complementary signature of the same underlying structure. This formulation suggests that while the underlying relaxation tensor is fundamental, what can be experimentally observed is not solely a property of state, but of the trajectory taken through state space.

## Theory

### *The covariant relaxation tensor*

We develop this framework by first expressing dissipation as a frame-independent object in Liouville space, then projecting onto observables to recover the Bloch picture, and finally by showing how noncommuting control reveals the tensor structure experimentally. To represent dissipation in a form that is independent of measurement frame and control protocol, we begin with the general equation of motion for the density operator,

$$\dot{\rho} = -\frac{i}{\hbar}[\hat{\mathcal{H}}(r), \rho] + \mathcal{D}(\rho), \qquad (1)$$

where $\hat{\mathcal{H}}(r)$ is the system's Hamiltonian, which may be spatially dependent, and $\mathcal{D}(\rho)$ describes irreversible dynamics arising from coupling to the environment. For Markovian open quantum systems, the dissipator may be expressed in GKSL form.[11,14,17–20] Expanding in a Hermitian operator basis $\{F_\alpha\}$, one obtains

$$\mathcal{D}(\rho) = -\frac{1}{2}\sum_{\alpha\beta} R_{\alpha\beta}\, [F_\alpha, [F_\beta, \rho]], \qquad (2)$$

where $R_{\alpha\beta}$ is real, symmetric, and positive semidefinite. Here $\alpha$ and $\beta$ label directions in operator space, and the double commutator describes the successive action of $F_\beta$ and $F_\alpha$ on $\rho$.[22] A full derivation of Eq. (2) is provided in §§2.1–2.5 of the Supplementary Text.

In conventional formulations, $R_{\alpha\beta}$ is defined with respect to a chosen operator basis and is therefore representation dependent.[2,13,23] Thus, the same physical dissipative process is described by different matrices under changes of basis or frame; this dependence becomes pronounced in driven systems where the effective Hamiltonian and measurement basis vary in time and space (vide infra). Here we reinterpret $R_{\alpha\beta}$ as the coordinate representation of an underlying bilinear form – a map that takes pairs of operator components to a scalar decay rate, linearly in each argument, independently of the basis used to represent them. We refer to this object as the covariant relaxation tensor $R_{\mu\nu}$. Its components transform covariantly under changes of operator basis, while the tensor itself is basis-independent.[24–27]

This distinction is the central claim of the framework: $R_{\mu\nu}$ is the fundamental object governing dissipation; what any experiment accesses is a control-dependent projection of it. The implications of this claim are developed in the two subsections that follow.

***Projection onto observables: recovering the Bloch picture***

Experimental measurements probe specific projections of $R_{\mu\nu}$ rather than the tensor itself. To connect the Liouville space picture to observable dynamics, we project onto a set of expectation values $M_i = \mathrm{Tr}(\rho F_i)$ , where $\{F_i\}$ is a chosen operator basis such as the Cartesian spin operators.[28] The evolution of $M_i$ follows from Eq. (1) and takes the reduced form

$$\dot{M}_i = \sum_j L_{ij} M_j + b_i, \qquad (3)$$

where $L_{ij}$ is the generator of linear dynamics in this representation and $b_i$ sets the equilibrium magnetization. Decomposing into coherent and dissipative parts,

$$L_{ij} = \Omega_{ij} - R_{ij}, \qquad (4)$$

where $\Omega_{ij}$ is antisymmetric and describes precession, and $R_{ij}$ is symmetric and describes relaxation. The components $R_{ij}$ are projections of the underlying tensor $R_{\mu\nu}$ onto the chosen observable basis; further derivation is in §2.3 of the Supplementary Text.

The standard Bloch picture corresponds to the special case in which this projected matrix is diagonal.[1,3,29–31] In a basis aligned with the principal relaxation axes,

$$R_{zz} = R_1, \qquad (5)$$

$$R_{xx} = R_{yy} = R_2, \qquad (6)$$

with off-diagonal elements assumed to vanish, and longitudinal and transverse processes appear decoupled. In Redfield theory, however, these rates arise from the same spectral densities of environmental fluctuations, which signals their common origin.[13,32] In the present framework, this diagonal form is understood as a consequence of choosing a basis in which $R_{\mu\nu}$ happens to project diagonally and is not a fundamental separation of processes.

More generally, off-diagonal components of $R_{ij}$ need not vanish. They encode coupling between different components of the observable vector, consistent with their shared origin in the environmental spectral densities (SI §2.8).[13,32] $R_1$ and $R_2$ are therefore not fundamental parameters but complementary projections of $R_{\mu\nu}$ onto a specific measurement basis. Different bases, or different effective frames induced by control, yield different representations $R_{ij}$ even when the underlying dissipative process is unchanged.

This basis-dependence is already well established in magnetic resonance. In the static limit, $R_{1\rho}$ relaxation measures decay along an axis set by pulse control rather than by the static field.[4] During pulsed spin locking, an applied field of amplitude $\omega_1$ at offset $\Delta\omega$ produces an effective field $\omega_e = \sqrt{\Delta\omega^2 + \omega_1^2}$ tilted from $\hat{z}$ by $\theta = \arctan(\omega_1/\Delta\omega)$ . In the $\omega_e\tau_c \ll 1$ limit, magnetization locked along $\hat{n} = (\sin\theta, 0, \cos\theta)$ decays at $R_{1\rho} = R_1\cos^2\theta + R_2\sin^2\theta$.[33,34] This is the quadratic form $\hat{n}^T R\, \hat{n}$ evaluated along $\hat{n}$, which is a diagonal projection of the underlying relaxation tensor onto the measurement frame. This projection behavior is generalized in the tensor framework, and the case of $R_{1\rho}$ is illustrative because it contains two implicit conditions. First, the longitudinal–transverse off-diagonal components $R_{xz}$ and $R_{yz}$ are assumed to vanish. This condition holds when the relaxation-inducing fluctuations possess axial symmetry around $\hat{z}$. Second, the spectral density that is sampled along the spin-lock direction is evaluated at zero frequency rather than at $\omega_e$. Taking into account this distinction gives the modified expression, $R_{1\rho} = R_1\cos^2\theta + R_2\sin^2\theta + \sin^2\theta\,[J_{zz}(\omega_e) - J_{zz}(0)]$, which is the origin of rotating frame relaxation dispersion.[5,35] Chemical exchange can also contribute through this term with the exchange correlation time substituted for $\tau_c$ and arises from a change in the spectral density sampled along a given direction, which is a different mechanism than nonzero off-diagonal structure in $R_{\mu\nu}$. The location of the tensor projection relative to the second-order step in Redfield theory determines the allowed angular structure, as illustrated by the contrast with orientation-dependent $R_2$ in the Discussion and SI §2.8.

### *Noncommuting control as a probe of tensor structure*

The preceding analysis shows that the measured relaxation matrix depends on the choice of observable basis. In the case of $R_{1\rho}$, this basis is a single fixed axis set by the effective field. As discussed above, projection onto a fixed axis yields the quadratic form along one direction. By design, such a measurement cannot report off-diagonal tensor components, since they require the projection to vary. Under driven evolution more generally, however, the effective basis is not fixed – it is determined by the full time-ordered history of the control sequence. This is the mechanism by which the tensor structure of $R_{\mu\nu}$ becomes experimentally accessible. When a system evolves under a time-dependent Liouvillian $\mathcal{L}(t)$, the formal solution is

$$\rho(t) = \mathcal{T}\exp\left(\int_0^t \mathcal{L}\,(t')\,dt'\right)\rho(0), \qquad (7)$$

where $\mathcal{T}$ denotes time ordering. When $\mathcal{L}(t_1)$ and $\mathcal{L}(t_2)$ do not commute at different times, the evolution depends on the ordering of the applied transformations. (See §3 of the Supplementary Text for a derivation and examples.) For periodically driven systems, the stroboscopic dynamics may be written in terms of an effective Floquet generator $\mathcal{L}_{\text{eff}}$ satisfying

$$\rho(nT) = \exp(nT\mathcal{L}_{\text{eff}})\,\rho(0), \qquad (8)$$

where $n$ is the cycle index and $T$ the cycle period. When instantaneous generators do not commute, $\mathcal{L}_{\text{eff}}$ cannot be obtained by simple averaging but instead incorporates commutator terms through a Magnus expansion [36],

$$\mathcal{L}_{\text{eff}} = \frac{1}{t}\int_0^t \mathcal{L}(\,t_1)\,dt_1 + \frac{1}{2t}\int_0^t dt_1 \int_0^{t_1}[\mathcal{L}(t_1), \mathcal{L}(t_2)]dt_2 + \cdots\,, \qquad (9)$$

where the commutator terms encode the ordering dependence of the dynamics. When generators commute, the series truncates at first order and the effective relaxation matrix reduces to the fixed-frame projection; noncommuting sequences generate the higher-order terms that expose off-diagonal structure. There is a geometric meaning of this ordering dependence. In general, transporting along one generator and then a second does not yield the same result as operations applied in the reverse order. Failure of these paths to close is measured by their commutator, which is classically an indicator of underlying curvature.[37] This structure also applies in Liouville space, where superoperators act as vector fields on the density $\rho$. In this case, the closure defect appears as a double commutator.[22] A pulse sequence creates a path through the space of control parameters that can be engineered into the Hamiltonian, and the associated curvature is of a connection defined over the control parameters. The covariant formulation of this is developed in Supplementary Text §§2.2 and 2.11, and the geometric meaning is illustrated in Fig. 1.

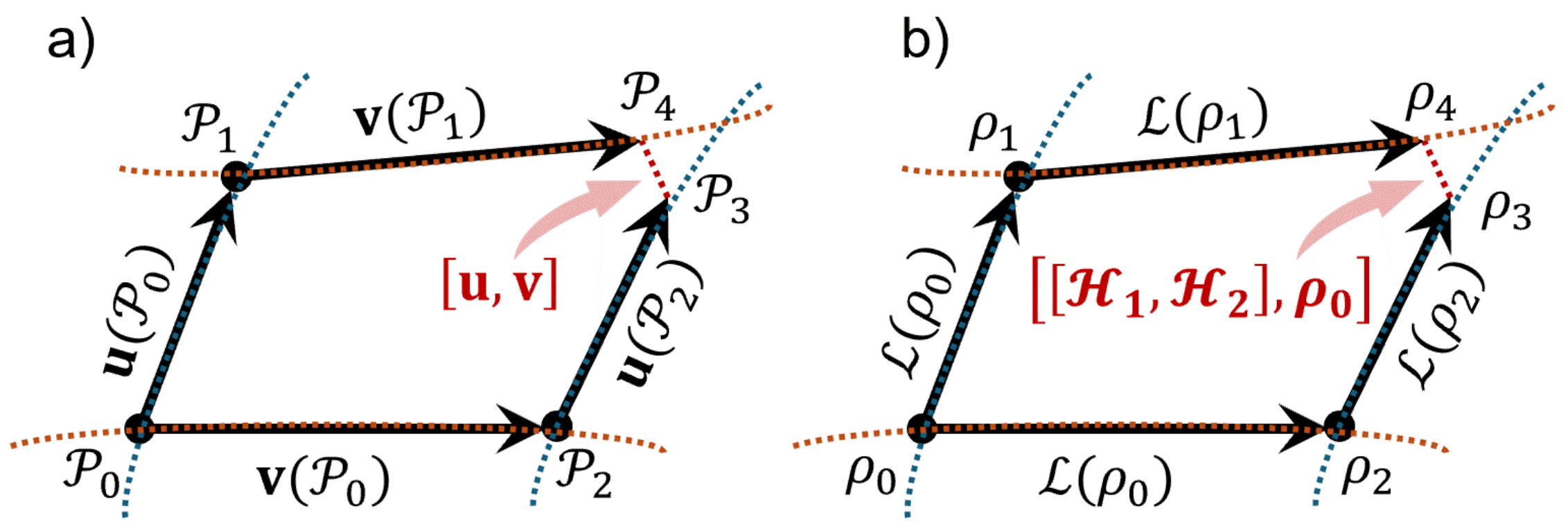

**Fig. 1. Commutators as geometric transport. (a)** In general, for two vector fields $u$ and $v$, transporting a point first along $u$ and then $v$ does not necessarily return to the same location as transporting in the opposite order. The resulting failure of the quadrilateral to close defines the commutator $[u, v]$, providing a geometric measure of curvature. **(b)** The same construction applies in Liouville space, where superoperators act as vector fields on the density operator $\rho$. The closure defect becomes a double commutator, linking noncommuting generators to curvature-like evolution. Relaxation shrinks this loop, representing entropy production and energy exchange. Further mathematical description is provided in §2.11 of the Supplementary Text.

Equation (2) contracts successive adjoint actions with $R_{\alpha\beta}$ to give dissipation. Since $R_{\alpha\beta}$ is symmetric, the antisymmetric difference between the two orderings cancels and only the symmetric part contributes. Because curvature is contained in the antisymmetric part, any measurable ordering dependence must arise from commutators of the full Liouvillian in Eq. (9). As shown in Supplementary Text §3.3, these include mixed terms such as $[\mathcal{L}_H(t_1), \mathcal{D}(t_2)]$, which couple the coherent and dissipative parts of the generator and contribute to the measured relaxation matrix. Together, pulse control and anisotropy reshuffle which combinations of observables decay independently and which become coupled.

Projecting onto observables yields an effective relaxation matrix $R_{ij}^{(\mathrm{eff})}$ defined by the dissipative part of $\mathcal{L}_{\mathrm{eff}}$. This is the quantity accessed experimentally. Crucially, $R_{ij}^{(\mathrm{eff})}$ is not, in general, related to a single representation of $R_{\mu\nu}$ by a change of basis. Rather, it is a control-dependent projection of the underlying tensor that incorporates the effects of noncommutation and time ordering. Pulse sequences that traverse a single direction in operator space – e.g., single-axis driving, with time-independent Hamiltonians – sample a restricted projection and yield effectively diagonal $R_{ij}^{(\mathrm{eff})}$ .[38,39] Sequences that traverse multiple noncommuting directions mix components of $R_{\mu\nu}$ through the commutator structure of $\mathcal{L}_{\text{eff}}$, exposing off-diagonal elements that a fixed-frame description would miss.

In systems with spatially dependent or anisotropic Hamiltonians, the local Liouvillian varies from point to point in the sample, and this noncommutation also appears as curvature of the superoperator field (Fig. 2), with additional commutator terms that mix coherent and dissipative contributions to the effective dynamics.[11,40] In such cases, a single matrix representation does not provide a consistent description across frames, whereas the underlying tensor structure remains well-defined. In general, realistic materials with spatial heterogeneity in their Hamiltonians and spectral densities exhibit spatial variation in the local Liouvillian generator. This spatial structure, involving nonuniform decay rates and principal axes, can produce measurable effects. The symmetry structures underlying this behavior, including continuous symmetries as generators commuting with the Liouvillian, and symmetry breaking as the failure of $R_{\mu\nu}$ to conserve invariant subspaces, are developed formally in §3 of the Supplementary Text.

### *Experimental predictions*

Pulse sequences that apply successive rotations along noncommuting axes should produce: (i) transverse magnetization trajectories that deviate from monotonic contraction along a fixed axis, reflecting ordering-dependent mixing; (ii) effective relaxation matrices with non-negligible off-diagonal components that vary with control parameters such as transmitter frequency offset; and (iii) geometric phase accumulation under cyclic sequences, arising from the holonomy of noncommuting generators and separable from dissipative amplitude decay.[41–44] The following section demonstrates each of these signatures in hyperpolarized $^{13}C$ spins in diamond with NV centers.

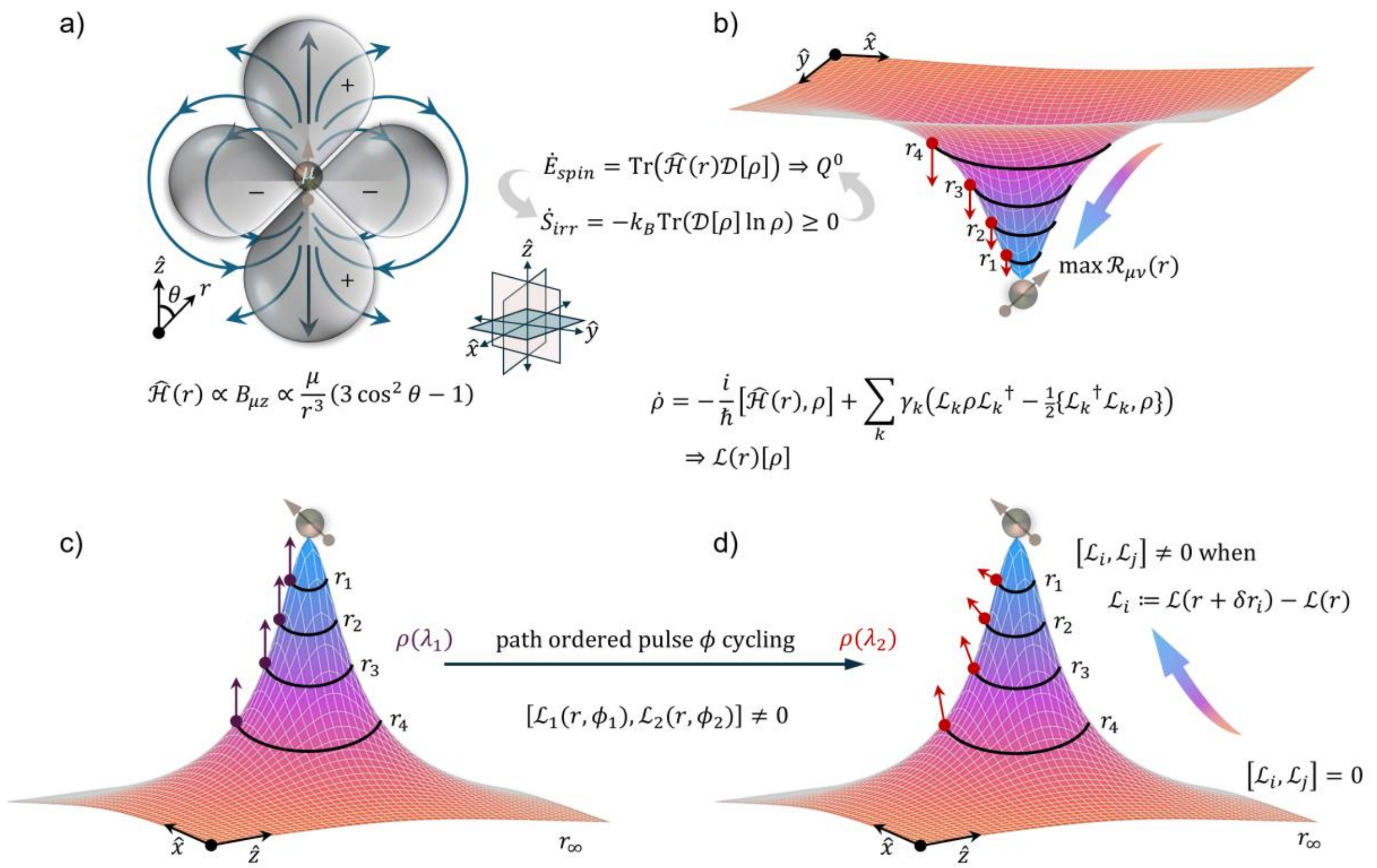

**Fig. 2. Field of local superoperators whose curvature is sourced by spatially variant relaxation dissipators and anisotropic dipolar Hamiltonians. (a)** Dipolar Hamiltonian modeled by a central unpaired electron (e.g., a paramagnetic center or nitrogen vacancy) whose radial and angular dependence is given by $\hat{\mathcal{H}}(r) \propto \frac{\mu}{r^3}(3\cos^2\theta - 1)$. **(b)** An $x, y$ embedding diagram (a 2D projection of a higher-dimensional curved surface) helps to visualize curvature of this field. An anisotropic static Hamiltonian leads to radially dependent relaxation rates that increase toward the dipolar source. The resulting spatial dependence of dissipators, written here in Lindblad form, is described by a sum over $k$ components weighted by a material-dependent coefficient $\gamma_k$. Together, these define a field of local Liouville generators $\mathcal{L}(r)$ whose noncommutativity sources the curvature depicted. **(c)** A $z, x$ embedding diagram of curvature in this field for an initial time-like slice of spin density parameterized by $\lambda_1$. **(d)** The $z, x$ embedding diagram of spin density at later time parameterized by $\lambda_2$, following cyclic pulse phase transformations. Here, spin isochromats depicted at arbitrary radii $r_1$, $r_2$, $r_3$, and $r_4$ experience both (i) varied rotation (manifested as holonomy) due to noncommutativity of the pulse train and the spatial anisotropy of the Hamiltonian via $[\mathcal{L}_i(r_i, \phi_i), \mathcal{L}_j(r_j, \phi_j)] \neq 0$ as well as (ii) varied contraction (shrinking) of projected vector components due to spatially dependent relaxation following a pulse train.

## Results

### *Transverse trajectories as a diagnostic for generator commutation*

The structure of spin trajectories during a pulse train reflects the algebra of the generators that define the sequence. When the effective dynamics are dominated by a single generator, successive pulses commute and transverse magnetization evolves along smooth trajectories in the $S_x$–$S_y$ plane.[45–47] This behavior is observed for pulse trains consisting of an initial pulse along $y$ followed by repeated pulses along $x$ (Fig. 3a). In this case, the trajectories contract smoothly toward the origin under relaxation, with larger per-pulse rotations at longer pulse widths.

When successive pulses act along noncommuting directions, the evolution cannot be reduced to a single effective generator. For sequences that cyclically traverse orthogonal axes $(x, y, -x, -y)$, each segment applies a rotation about a different generator, and the resulting dynamics depend on the ordering of these operations. The trajectories in Fig. 3b form multi-lobed, pinwheel patterns because successive pulses act about different axes, moving the transverse magnetization into different regions of the $S_x$–$S_y$ plane. Here, commutators between successive generators introduce effective terms that are not present in any individual segment of the sequence (SI §5.1). Whereas commuting sequences produce simple contraction along a fixed direction, noncommuting sequences generate winding trajectories that sample multiple directions in spin space.

The observed trajectories are shaped by both control and dissipation. Subtle distortions of the idealized patterns, such as rounding of corners and clustering of points near the $\pm x$ and $\pm y$ crossings (Fig. 3b), are consistent with relaxation acting anisotropically in state space (further discussion is provided in SI §6.5). The clustering is consistent with the trajectory slowing where it is locally aligned with an eigenvector of $R_{\mu\nu}$. In the language of the framework developed above, these are the directions along which the projected relaxation matrix $R_{ij}$ has its largest eigenvalues. This directional sensitivity of dissipation is an experimental imprint of the tensor structure of $R_{\mu\nu}$, and motivates the quantitative extraction carried out in the following section.

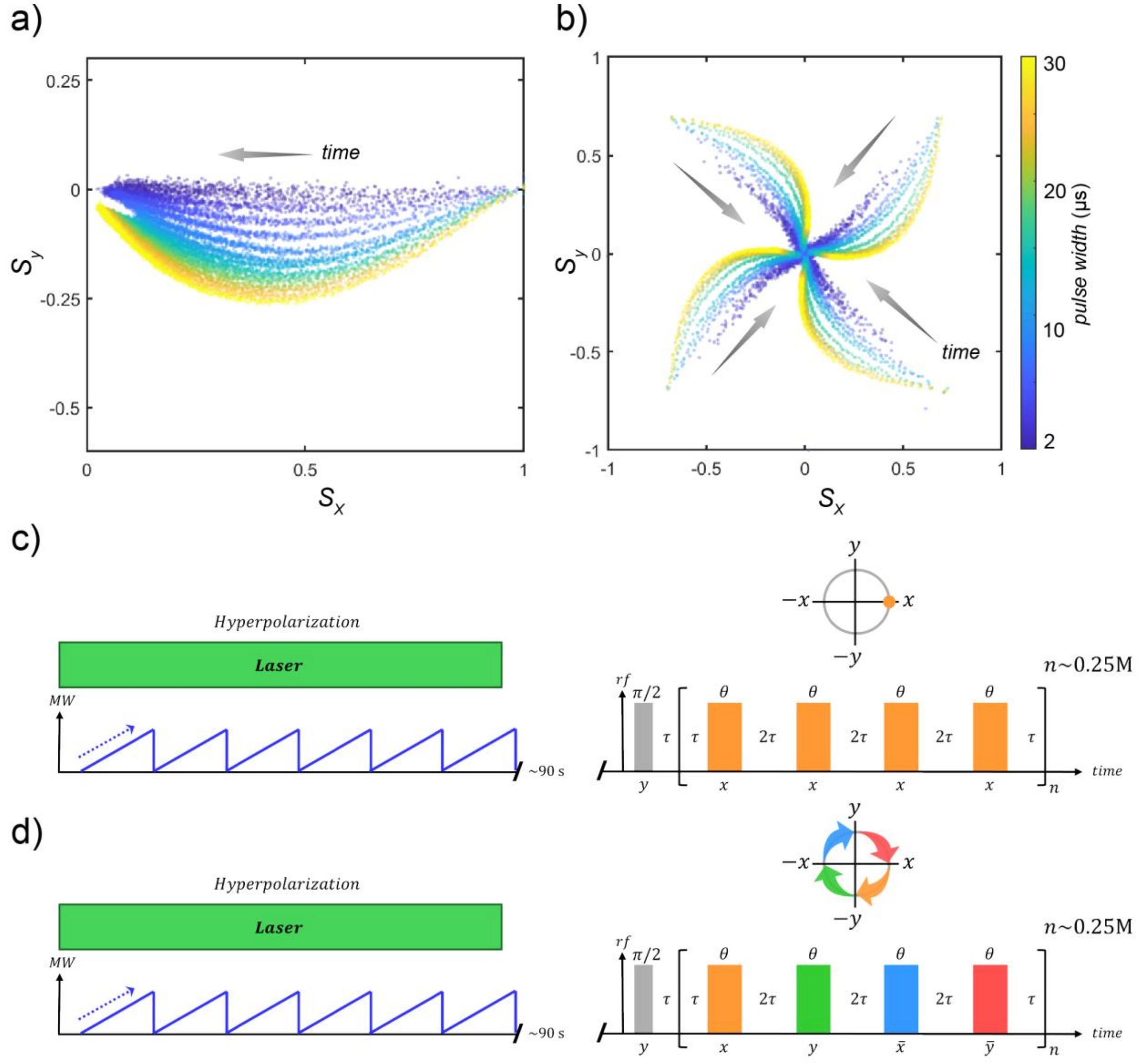

**Fig. 3. Spin trajectories reveal generator commutation. (a)** Raw transverse plane trajectory for pulse trains with effectively commuting generators, consisting of an initial pulse along $y$ followed by repeated pulses along $x$ for $^{13}C$ nuclear spins in diamond with nitrogen-vacancy (NV) centers. Trajectories are shown for pulse widths ranging from 2 μs (purple) to 30 μs (yellow). The effect of transmitter frequency offsets from −5 to +5 kHz is shown in the Supplementary Materials. **(b)** Raw transverse plane trajectory for noncommuting pulse trains, consisting of an initial $\pi/2$ pulse along $y$ followed by a repeating $(x, y, -x, -y)$ sequence. Pulse widths again range from $\theta = 2$ μs (purple) to 30 μs (yellow). The resulting trajectories reflect ordering-dependent evolution arising from noncommuting generators. In both plots, gray arrows indicate the direction associated with time evolution as progression toward the origin. **(c)–(d)** Pulse sequences corresponding to the measurements in **(a)** and **(b)**, respectively.

***Relaxation tensor tomography across control space***

Transmitter frequency offset $\Delta\omega$ provides an expedient control parameter for probing $R_{\mu\nu}$ systematically. On resonance, a pulse train rotates magnetization about axes fixed in the transverse plane. Introducing a finite $\Delta\omega$ adds a $z$-axis precession term $\widehat{\mathcal{H}}_{\text{offset}} = -\Delta\omega S_z$ between pulses; since $\widehat{\mathcal{H}}_{\text{offset}}$ does not commute with the transverse pulse Hamiltonians $\widehat{\mathcal{H}}_{\text{rf}}(t)$, this mixes operator space directions and reveals the off-diagonal structure of $R$. Varying $\Delta\omega$ therefore traces a controlled path through the space of tensor projections.

The off-diagonal components depend linearly on the offset. In the rotating frame under the rotating wave approximation,

$$\widehat{\mathcal{H}}_{Rot}(t) = -\Delta\omega S_z + \omega_1(t)\left[\cos\phi(t)\, S_x + \sin\phi(t)\, S_y\right], \qquad (10)$$

where $\Delta\omega = \omega_0 - \omega_{\text{rf}}$ is the offset frequency, $\omega_1$ the rf amplitude, and $\phi(t)$ the pulse phase. (The derivation from the laboratory frame is in Methods.) For zero-winding trains, $\widehat{\mathcal{H}}_{\text{rf}}(t) \propto S_x$ at all times. For phase-wound trains, the pulse axis cycles through noncollinear generators. The effective propagator over one cycle follows from a Magnus expansion,

$$U(T) = \exp\left(-i(\Omega_1 + \Omega_2 + \cdots)\right), \qquad (11)$$

where $\Omega_1 = \int_0^T \widehat{\mathcal{H}}(t)dt$, $\Omega_2 = -\frac{i}{2}\int_0^T dt_1 \int_0^{t_1}\left[\widehat{\mathcal{H}}(t_1), \widehat{\mathcal{H}}(t_2)\right]dt_2$, and so on. The second order term introduces the key commutators $\left[S_x, S_y\right] = iS_z$, $[S_x, S_z] = -iS_y$, and $\left[S_y, S_z\right] = iS_x$. Because $\widehat{\mathcal{H}}_{\text{offset}}$ does not commute with any $x$ or $y$ pulse, the second-order term contains contributions proportional to $\left[\widehat{\mathcal{H}}_{\text{rf}}, \widehat{\mathcal{H}}_{\text{offset}}\right] \propto \Delta\omega$ in all sequences. In phase-wound pulse trains, additional ordered combinations of $S_x$, $S_y$, and $S_z$ appear because alternating $x$ and $y$ pulses do not commute with each other. The result is that the off-diagonal effective relaxation components vary linearly in $\Delta\omega$ rather than as an even function.

Relaxation tensor fitting is an operator identification problem. To extract the tensor components quantitatively, we fit a linear generator to the stroboscopic transverse evolution. From $S_x(n)$ and $S_y(n)$ for pulse index $n$, we solve for the 2×2 effective generator $A_{\text{eff}}$ via

$$\frac{d}{dn}\begin{pmatrix} S_x \\ S_y \end{pmatrix} = A_{\text{eff}}\begin{pmatrix} S_x \\ S_y \end{pmatrix}, \qquad (12)$$

and decompose into symmetric (dissipative) and antisymmetric (precession) parts,

$$A_{\text{eff}} = -R_{\perp} + \Omega, \qquad (13)$$

where $\Omega = \frac{1}{2}(A_{\mathrm{eff}} - A_{\mathrm{eff}}^T)$ generates precession in the transverse plane and $R_\perp = -\frac{1}{2}(A_{\mathrm{eff}} + A_{\mathrm{eff}}^T)$ is the transverse projection of the effective relaxation tensor. The components $R_{11}$ and $R_{22}$ represent decay rates along $x$ and $y$; $R_{12}$ and $R_{21}$ represent cross-coupling between them, rotating the principal relaxation axes off $x$ and $y$. Since we expect strong $\Delta\omega$-dependence but not necessarily pulse width $\tau_p$ dependence, we average over $\tau_p$ for each offset,

$$R_{ij}(\Delta\omega) = \langle R_{ij}(\Delta\omega, \tau_p) \rangle_{\tau_p}. \qquad (14)$$

Figure 4 shows the resulting components for commuting and noncommuting sequences. For the commuting (zero-winding) train, $R_\perp$ remains approximately diagonal across the full range of offsets explored, consistent with dynamics dominated by a single effective generator, for which the tensor projects onto a fixed frame. For the noncommuting (phase-wound) trains, the picture changes markedly. The off-diagonal components $R_{12}$ and $R_{21}$ become non-negligible and reach ≈0.1 $s^{-1}$ at ±5 kHz offset (Figs. 4b and 4c), roughly a third to a half of the diagonal rates $R_{11}$, $R_{22}$ ≈0.15–0.3 $s^{-1}$ (Figs. 4a and 4d), and vary approximately linearly with $\Delta\omega$, with opposite slopes for positive and negative winding. The zero-winding control remains near zero (≲0.01 $s^{-1}$) across the full offset range. This demonstrates the linear-in-$\Delta\omega$ behavior predicted by the Magnus expansion above. Changing the sign of the offset changes the sign of the ordered correction terms; the phase-wound pulse train produces a response of the form $R_{12}^{(\mathrm{eff})}(\Delta\omega) \approx a + b\Delta\omega + \cdots$, where $a \sim R_{12}^{(\mathrm{zero-winding})}$. Similar behavior occurs for $R_{21}^{(\mathrm{eff})}(\Delta\omega)$, with opposite sign. The diagonal components $R_{11}$ and $R_{22}$ also shift between commuting and noncommuting conditions, reflecting a redistribution as the principal dissipative axes rotate away from the laboratory frame $x$ and $y$ directions. Further description of fitting is described in §6 of the Supplementary Text.

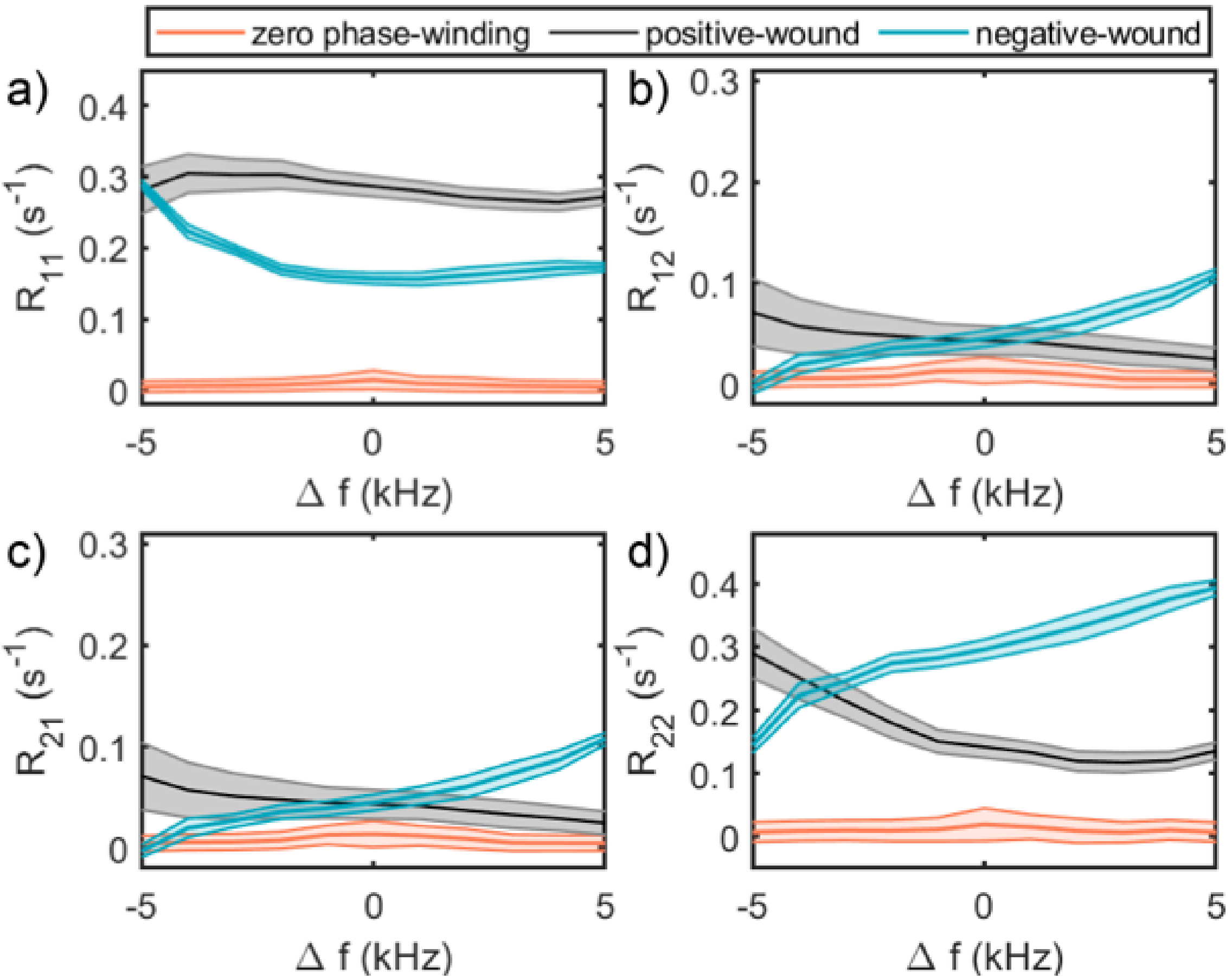

**Fig. 4. Effect of generator commutation on projected relaxation tensor components.** For $^{13}$C nuclear spins in diamond with NV centers, **(a)** $R_{11}$, **(b)** $R_{12}$, **(c)** $R_{21}$, and **(d)** $R_{22}$ are shown for commuting (zero-winding, red) and noncommuting (positive and negative phase-wound, in gray and blue, respectively) pulse trains as a function of transmitter offset frequency. These measured $R$ represent effective transverse dissipative projections of the Floquet generator. The shaded regions represent standard error, and fits were averaged across pulse widths from 2 to 30 μs as reported in Eq. (14).

These results are the central experimental finding of this paper. The off-diagonal components $R_{12}$ and $R_{21}$ observed under noncommuting control are the experimental signature of the tensor structure of $R_{\mu\nu}$. They arise because the phase-wound sequences sample projections of the underlying tensor that are inaccessible to a single-axis sequence, mixing components through the commutator structure of $\mathcal{L}_{\text{eff}}$ as described in the Theory section. A scalar or diagonal relaxation model cannot account for their magnitude, sign structure, or systematic dependence on $\Delta\omega$. Extracting the full tensor, including $R_{33}$ and the $z$-mixed transverse components, would probe the coupling between coherence loss and energy exchange. Such measurements will require modified sequences and are the subject of ongoing work.

***Holonomy as an independent confirmation of noncommuting dynamics***

The tensor tomography of the preceding section probes the dissipative structure of $R_{\mu\nu}$ directly. A complementary and independent signature of noncommuting dynamics is provided by the accumulation of geometric phase under cyclic pulse sequences, which is insensitive to dissipation. For the phase-wound pulse trains – the same noncommuting sequences whose trajectories appear in Fig. 3b and whose tensor projections were extracted in Fig. 4 – the transverse trajectories trace rotating nested-square patterns whose piecewise evolution is generated by successive pulses (Fig. 5a). Introducing a transmitter frequency offset causes an additional slow rotation to be superimposed on this pattern (Fig. 5b).

This additional rotation in Fig. 5b arises from the $-\Delta\omega S_z$ term introduced by the frequency offset in the Hamiltonian of Eq. (10). Through the second order terms of the Magnus expansion in Eq. (11), the commutators $[S_z, S_x] = iS_y$ and $[S_z, S_y] = -iS_x$ generate an in-plane precession that grows linearly with $\Delta\omega$. The closed-loop holonomy of the four-pulse cycle is a separate quantity that accumulates a fixed phase per cycle, independent of offset. This quantity is the geometric phase measured interferometrically in Fig. 5c. Geometric phase has been observed previously across a wide range of spin systems and field conditions, and its invariance to dynamic phase contributions is well established.[41,44,48–51]

The geometric phase here is extracted interferometrically by taking the half-difference of complex echo amplitudes between positive- and negative-winding sequences, referenced to the zero-winding control.[52] This isolates the odd-in-winding geometric component while cancelling dynamic phase contributions to first order. As shown in Fig. 5c, the accumulated phase increases linearly with cycle number and is symmetric in the sign of the winding. Here the accumulation persists robustly across nearly one million rf pulses, remaining insensitive to transmitter frequency offset and relaxation-induced amplitude decay until there is no detectable signal. This reflects the fact that geometric phase and dissipative dynamics obey different algebras and evolve independently. (This separation is proven in §5.6 of the Supplementary Text and verified empirically in Fig. S1.)

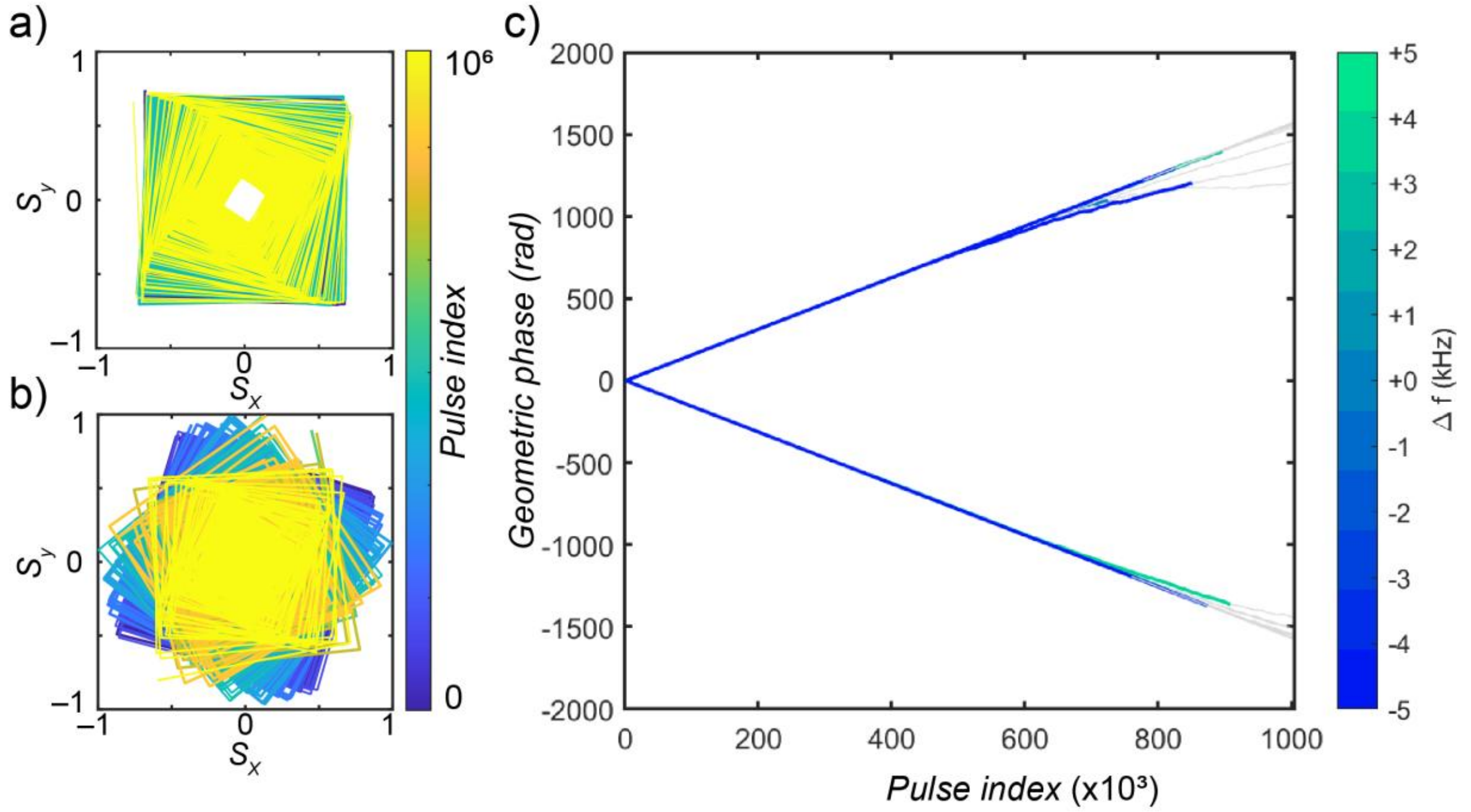

**Fig. 5. Holonomy and parallel transport in spin dynamics. (a)** Continuous transverse plane trajectory corresponding to resonant excitation as shown in Fig. 3b. **(b)** The effect of introducing a 4 kHz transmitter frequency offset on the pulse train shown in **(a)** with continuous positive phase winding of $(x, y, -x, -y)$. **(c)** Geometric phase extracted interferometrically from positive and negative phase winding experiments, referenced to a zero winding (commuting, all $x$ pulses), for transmitter offsets ranging from $-5$ kHz to $+5$ kHz in 1 kHz steps. Notably, geometric phase is robust to the effects of dynamic phase and relaxation, and continues accumulating identically until degradation of SNR (gray traces in **(c)**, where the bulk signal falls below the noise floor and coherent phase accumulation is lost). Pulse trains consisted of variable echo times and number of pulses (approximately 500,000 to >1 million; details in Methods) and were downsampled to 1000 points to compare uniformly.

The tensor tomography of Fig. 4 and the holonomy of Fig. 5 probe the same noncommuting dynamics through different lenses. Figure 4 reveals how $R_{\mu\nu}$ is sampled through control-dependent trajectories; this is a dissipative measurement that is frame- and sequence-dependent. Figure 5 reveals the holonomy of those trajectories, which is a geometric quantity that is insensitive to dissipation and robust across more than a million pulses. Their agreement across independent experimental signatures strengthens the interpretation of each.

## Discussion

The central result of this work is that spin relaxation, conventionally parameterized by two independent scalar rates $R_1$ and $R_2$, is well described as a covariant tensor $R_{\mu\nu}$ whose projections depend on the measurement frame and the full time-ordered history of the control sequence. The conventional rates are not fundamental parameters, but a representation-dependent projection that is accurate under single-axis, time-independent evolution. This approximation breaks down when pulse control traverses noncommuting directions in operator space. The breakdown is not a failure of the Bloch or Redfield pictures so much as a signal that the scalar picture is a projection of a richer structure.

Another well-known case where this structure appears is the orientation dependence of $R_2$ in ordered solids.[53–58] Such $R_2$ anisotropy results from the same algebraic mechanism applied to a different object in a different space. In this case, the projected object is the dipolar coupling tensor $D$, whose indices label directions in the sample. Projection onto a measurement frame gives the secular dipolar frequency as $\omega_D = \hat{B}^T \cdot D \cdot \hat{B} = -2b\, P_2(\cos\beta)$, with $P_2$ the second Legendre polynomial, $\beta$ the angle between the internuclear vector and the field, and $b$ the dipolar coupling constant. As molecular motion reorients this three-dimensional dipolar-coupling pattern relative to the measurement frame, $R_2$ samples the frequency spectrum of the resulting modulation.

In the case of $R_{1\rho}$, the relaxation tensor is projected by a continuous spin-lock onto one static, tilted axis in the rotating frame. In cases with driven parallel transport, such as the phase-incremented pulse trains shown in this work and in Ref. [52], the relaxation tensor is projected onto the toggling frame where the interaction frame changes discretely. For $R_2$ anisotropy, the dipolar coupling tensor is first projected from a crystal frame onto the laboratory frame (in pulsed spectroscopy; for continuous wave measurements this is the rotating frame). This contraction retains only the secular component $D_{zz}$ along $B_0$, which is modulated by molecular motion. The correlation function of these fluctuations yields the spectral densities, and the Redfield rates are proportional to the square of this angular factor, $R_2 \propto \langle \delta\omega_D^2 \rangle \tau_c$. Therefore, $D$ maps to $R_{\mu\nu}$ through the second moment of the projected coupling. In contrast, for $R_{1\rho}$, the relaxation tensor is already second order in the coupling, so the contraction onto the effective field axis is applied once and its angular dependence contains Legendre polynomials only of orders 0 and 2 (via the decomposition $\frac{R_1+2R_2}{3} P_0 + \frac{2}{3}(R_1 - R_2)P_2(\cos\theta)$, as shown in SI §2.8). With orientation-dependent $R_2$, the rate is quadratic in the projected coupling, extending its angular dependence to order 4 (as $[P_2(\cos\beta)]^2 = \frac{1}{5}P_0 + \frac{2}{7}P_2 + \frac{18}{35}P_4$, shown in SI §2.8). The difference here is in the order of operations. For orientation-dependent $R_2$, the dipolar coupling tensor is projected first and the rate is then formed from the second moment of the projected coupling, allowing angular terms through $P_4$. For $R_{1\rho}$, the projected quantity is the second-order relaxation tensor itself, yielding only $P_0$ and $P_2$ contributions.

The tensor structure reframes the boundary between coherent and irreversible dynamics. In the scalar $R_1, R_2$ picture, dephasing and energy exchange are treated as independent processes, and one might imagine refocusing the former while leaving the latter unchanged. The tensor framework shows this separation to be representation-dependent. Under noncommuting control, off-diagonal components of $R_{\mu\nu}$ couple transverse dephasing to longitudinal population transfer. This limit of reversibility, i.e. how much of the dissipated coherence is refocusable, depends on both the intrinsic algebra governing the system and the interaction with the environment encoded in $R_{\mu\nu}$. Measurements of the $z$-mixed components of the tensor could, in principle, probe this coupling between irreversible coherence loss and population transfer.

The holonomy measurements of Fig. 5 complement the tensor tomography of Fig. 4 by demonstrating that the noncommuting structure of the control generators leaves an imprint on the dynamics that is independent of dissipation. The geometric phase accumulates through the algebraic structure of the generator sequence and persists robustly across more than one million pulses, immune to relaxation-induced amplitude decay. Holonomy therefore provides an invariant diagnostic of noncommutation that persists even as dissipation degrades the signal amplitude. It integrates the solid angle enclosed by the control trajectory, analogous to a gyroscope recording net rotation.

The robustness of geometric phase to dissipation underlies the appeal of holonomic quantum computing, where logical operations are implemented as loops in parameter space that are insensitive to certain control errors associated with dynamical phase fluctuations.[59,60] In this setting, the language of geometrodynamics inherently describes spin dynamics under control, and geometric phase accumulation tabulates the curvature enclosed by the control loop. The measurements here realize this principle in a spin system with well-characterized dissipation. This suggests that pulse sequences designed to maximize holonomy while minimizing dissipative projection – for example, by traversing large loops in operator space while remaining close to eigenvectors of $R_{\mu\nu}$ – may inform the design of holonomic gates with reduced sensitivity to dissipation, though demonstrating a gate-level advantage would require fidelity benchmarking on a logical encoding.

Several extensions follow from the present work. Global Floquet inversion, such as fitting a single underlying $R_{\mu\nu}$ simultaneously to measurements from multiple pulse families, would provide a consistency check on the tensor description and a more complete characterization of the dissipative structure. Sequences designed to probe $z$-mixed components would extend the tomography to the full tensor and quantify the coupling between coherence and energy relaxation. Finally, the connection between the spatial dependence of $R_{\mu\nu}$ and the curvature of the superoperator field suggests that spatially resolved tensor tomography, combined with the holonomy measurements demonstrated here, could provide a new tool for imaging heterogeneous dissipation in materials with structural disorder or paramagnetic centers.

## Conclusions

The framework developed here joins Bloch, Redfield, and Lindblad descriptions within a common coordinate-independent formulation. Its natural domain of application includes driven systems where time-dependent control creates effective Hamiltonians that vary with the pulse sequence, anisotropic or spatially heterogeneous materials where the principal relaxation axes are not aligned with the laboratory frame, hyperpolarized systems prepared far from thermal equilibrium, and strongly driven or sensing protocols where the secular approximation may not hold.

What changes with this perspective is not the underlying physics but the language available to describe it. A coordinate-independent tensor formulation provides a framework to compare relaxation measurements made under different pulse sequences and effective frames, to ask what projections of $R_{\mu\nu}$ are accessible under a given control scheme, and to quantify the coupling between irreversible dephasing and energy exchange. These questions are difficult to pose, let alone answer, within a scalar $R_1/R_2$ framework. More broadly, the connection between tensor structure, geometric phase, and spatial heterogeneity in the superoperator field suggests a path toward spatially resolved relaxation tomography in materials with structural disorder or paramagnetic centers – a regime where conventional relaxation models are most strained and a richer description is most needed.

## Methods

***Sample.*** Experiments used a single-crystal diamond sample (3 × 3 × 0.3 mm) from Element Six containing $^{13}C$ spins at natural abundance (1.1%) and an $NV^-$ concentration of ≈1 ppm, with each NV center surrounded by ~$10^4$ $^{13}C$ nuclei. The sample also hosts ≳20 ppm of substitutional nitrogen impurities (P1 centers). The [100] face was oriented approximately parallel to $B_0$.

***Apparatus.*** Hyperpolarization and interrogation were performed using a cryogenic field-cycling instrument described previously[39,61,62] enabling shuttling between low field ($B_{pol}$ = 36 mT) for optical dynamic nuclear polarization (DNP) and high field ($B_0$ = 9.4 T) for $^{13}C$ interrogation. Experiments were performed at 100 K. Hyperpolarization used 532 nm laser illumination (5 W, Coherent) of NV centers combined with microwave chirps swept across the NV electron paramagnetic resonance bandwidth (25 MHz) at 200 Hz, generated by a Tabor Proteus arbitrary waveform transceiver. Following hyperpolarization, the sample was shuttled to high field via a belt-driven actuator (~90 s shuttling time), where rf pulses at the $^{13}C$ Larmor frequency (~100 MHz) were generated by a Proteus AWT unit and detected inductively in windows between pulses.

***Pulse train sequences.*** Three pulse train families were applied to the hyperpolarized $^{13}C$ nuclei: zero-winding, positive-winding, and negative-winding.[52] All trains began with a $\theta$ pulse along $y$ followed by a sequence of $\theta$ pulses with phase-cycled axes. The zero-winding sequence applied all subsequent pulses along $x$. Positive-winding cycled pulse phases through $(x,y,-x,-y)$, while negative-winding cycled through $(x,-y,-x,y)$. For each sequence, pulse widths were varied from 2 to 30 μs, transmitter frequency offsets from −5 to +5 kHz in 1 kHz steps, and approximately 500,000 to >1 million pulses were applied per train. The transverse magnetization $S_x(n)$ and $S_y(n)$ were sampled stroboscopically between pulses and downsampled to 1000 points for uniform comparison across pulse trains.

***Data acquisition.*** The induced $^{13}C$ Larmor precession signal was captured by the Tabor Proteus AWT and digitized at 1 ns intervals in acquisition windows between pulses, then decimated 64-fold for memory and processing efficiency. A fast Fourier transform extracted complex signal amplitude in each acquisition window, and phase unwrapping was used to recover continuous rotating-frame phase $\phi_R$. The pair $(S, \phi_R)$ formed the basis for all subsequent analysis.

***Geometric phase extraction.*** Geometric phase was obtained interferometrically following the protocol described previously.[52] The half-difference of complex echo amplitudes between positive- and negative-winding sequences, referenced to the zero-winding control, isolates the odd-in-winding geometric component while cancelling $B_0$ drift, receiver phase, and other dynamical phase contributions to first order.

***Data analysis.*** All data analysis was performed in MATLAB (MathWorks, Natick, MA). Effective transverse relaxation tensors were extracted by fitting linear generators to the stroboscopic transverse trajectories as described in §6 of the Supplementary Text.

***Rotating frame convention.*** In the laboratory frame, the Hamiltonian with a transmitter frequency offset is

$$\hat{\mathcal{H}}_{Lab}(t) = -\gamma B_0 S_z + 2\omega_1(t)\cos\big(\omega_{\mathrm{rf}}t + \phi(t)\big)\, S_x \tag{15}$$

or equivalently, decomposing the radiofrequency (rf) part into quadrature,

$$\hat{\mathcal{H}}_{Lab}(t) = -\omega_0 S_z + \omega_1(t)\big[\cos\big(\omega_{\mathrm{rf}}t + \phi(t)\big)\, S_x + \sin\big(\omega_{\mathrm{rf}}t + \phi(t)\big)\, S_y\big] \tag{16}$$

where $\omega_0$ is the Larmor precession frequency, $\omega_1$ is the rf amplitude, and $\omega_{\mathrm{rf}}$ the rf frequency. In the rotating frame, the dependence on offset frequency $\Delta\omega = \omega_0 - \omega_{\mathrm{rf}}$ becomes explicit. Transforming with $U(t) = \exp(i\omega_{\mathrm{rf}} t S_z)$,

$$\hat{\mathcal{H}}_{Rot}(t) = U\hat{\mathcal{H}}_{Lab}U^T - \omega_{\mathrm{rf}} S_z. \tag{17}$$

Under the rotating wave approximation, this becomes

$$\hat{\mathcal{H}}_{Rot}(t) = -\Delta\omega S_z + \omega_1(t)\big[\cos\phi(t)\, S_x + \sin\phi(t)\, S_y\big], \tag{18}$$

which is Eq. (10) in the Results section.

**Acknowledgements:** S.N.F. gratefully acknowledges support as a Pines Magnetic Resonance Center postdoctoral fellow, and K.A.H gratefully acknowledges support from an NSF Graduate Research Fellowship. We thank the anonymous referees for their comments and suggestions, which have helped to improve this work. We also acknowledge funding from ONR (N00014-20-1-2806), AFOSR YIP (FA9550-23-1-0106), and DOE BES (DE-SC0026041).

**Data Availability:** The data that support the findings of this study are available from the corresponding author upon reasonable request.

**Competing Interests:** The authors declare no competing interests.

# Supplemental Material for
# A geometric framework for spin relaxation

Sophia N. Fricke *et al.*
Corresponding author. Email: snfricke@berkeley.edu

**This PDF file includes:**

# 1 Origins of $R_1$ and $R_2$ interdependence

## 1.1 Spin relaxation as entropy–energy exchange

Conventional NMR relaxation theory begins with a useful separation of time scales: spin–spin and spin–lattice processes are typically treated as distinct phenomena with different physical origins.[1, 2] Transverse relaxation ($R_2$) describes the progressive loss of phase coherence within an ensemble of precessing spins. This dephasing is statistical in nature – individual spins retain their energy, but their phases disperse, leading to a decay of observable magnetization. Although often described as irreversible, the partial recovery of coherence in a Hahn echo shows that this loss reflects dispersal of phase information rather than immediate dissipation of energy. Accordingly, $R_2$ relaxation is commonly interpreted as an entropy-like scrambling of transverse phase – a decay of coherence without net energy exchange. By contrast, longitudinal relaxation ($R_1$) involves direct coupling to the environment and the exchange of energy with the lattice, driving the recovery of the equilibrium magnetization along the external field. Together, $R_2$ and $R_1$ distinguish two complementary mechanisms: redistribution of entropic phase information within the spin ensemble and transfer of energy between the spins and their surroundings.

The utility of this approach is most evident in the extraction of spectral density functions $J(0)$, $J(\omega)$, and $J(2\omega)$, which encode the underlying atomic and molecular dynamics. It is equally clear, however, that the two relaxation mechanisms are not truly independent. Phase coherence occurs only after a perturbation, such as a $\pi/2$ pulse, creates transverse magnetization, and the subsequent evolution necessarily takes place in a fluctuating environment that can exchange energy with the spins and drive realignment with the static field. Separating transverse and longitudinal relaxation into entropic and energy-transfer processes is therefore an approximation. This coupling is already evident in standard derivations of the relaxation master equation (see, for example, Chapter 11 of the text in Ref. [3]).

Redfield theory[4] makes this coupling explicit by expressing the fluctuating magnetic field as

$$\delta\mathbf{B}(t) = \left(\delta B_x(t), \delta B_y(t), \delta B_z(t)\right) \tag{S1}$$

corresponding to the Hamiltonian

$$H_1(t) = -\gamma \mathbf{S} \cdot \delta\mathbf{B}(t). \tag{S2}$$

The relaxation rates follow from the spectral densities $J_{ij}(\omega)$ of these fluctuations:

$$R_1 = \gamma^2 \left[J_{xx}(\omega_0) + J_{yy}(\omega_0)\right] \tag{S3}$$

$$R_2 = \frac{1}{2} R_1 + \gamma^2 J_{zz}(0) \tag{S4}$$

with

$$J_{ij}(\omega) = \int_0^\infty \left\langle \delta B_i(0) \delta B_j(t) \right\rangle e^{i\omega t} dt \tag{S5}$$

obtained from second-order perturbation theory.[4]

Several features of these expressions show that longitudinal and transverse relaxation arise from the same underlying fluctuations:

(i) Both $R_1$ and $R_2$ are generated by the same stochastic fields. The motions that randomize phase, including molecular tumbling, dipolar fluctuations, lattice vibrations, or exchange, simultaneously enable energy exchange with the bath.

(ii) The term often called "pure dephasing," $J_{zz}(0)$, requires a fluctuating Zeeman field. Any such fluctuation necessarily carries spectral weight near $\pm\omega_0$, opening a pathway for $R_1$ relaxation.

(iii) Consequently, transverse fluctuations produce both population transfer (spin flips, $R_1$) and phase dispersion ($R_2$), leading to the inequality $R_2 \geq \frac{1}{2}R_1$.

This inequality has a simple geometric interpretation. Longitudinal and transverse relaxation arise from different projections of the same fluctuating field, so $R_2$ must be at least half of $R_1$. Equality occurs only when those fluctuations are isotropic.

These observations motivate a geometric description of spin dynamics. The environmental fluctuations that disperse phase also exchange energy with the spin populations, so $R_1$ and $R_2$ are not independent processes but associated aspects of a single stochastic evolution. In this view, relaxation corresponds to transport of the spin state on the Bloch sphere, with longitudinal and transverse decay arising as different projections of the same motion.

This perspective invites a continuity description analogous to diffusion or charge flow, in which phase coherence and energy are gradually redistributed through the state space. Spin echoes separate the reversible component of this transport from the irreversible component. Refocusable contributions reflect deterministic evolution, while the remaining decay signals genuine entropy production, corresponding to information that has flowed into unobserved degrees of freedom.

Diffusion, dipolar couplings, and stochastic fields may therefore be understood as geometric distortions that redirect this flow. Relaxation can emerge from transport of information across the manifold of states, without any breakdown of the underlying unitary dynamics. In Section 2, we formalize this picture at three complementary levels.

# 2 Theoretical development: toward a geometric framework for spin relaxation

To work toward a general, geometric description of spin relaxation, it is beneficial to distinguish three levels of representation that are well established in magnetic resonance. At a microscopic, abstract level, relaxation appears as a superoperator acting on the density operator, which we denote $R_{\alpha\beta}$. At the level of observables, this reduces to the familiar Bloch relaxation matrix $R_{ij}$ governing the evolution of magnetization components. At a macroscopic level, dissipation can be expressed through continuity laws for energy and entropic information, where relaxation enters as an exchange or source term that we represent by a covariant tensor $R_{\mu\nu}$. Here, covariance ensures that $R$ is frame-independent; thus it is the fundamental object that acts as a bridge between bases.

Although these objects act in different spaces and use different mathematical languages, they describe the same underlying physics: exchange of energy with the environment and dispersal of phase coherence within the spin ensemble (depicted in Fig. 2 in the main text). In what follows, we show how these descriptions are related and can in fact be subsumed by a geometric approach, allowing reversible refocusing processes to be distinguished from irreversible entropy production.

## 2.1 The superoperator $R_{\alpha\beta}$: the microscopic, abstract object of operator fields

Starting with the general Lindblad equation, which is Markovian, trace-preserving, and has completely positive dynamics, the equation of motion for a density operator may be written,

$$\dot{\rho} = -\frac{i}{\hbar}[H, \rho] + \mathcal{D}[\rho] \tag{S6}$$

The full Liovillian $L[\rho] \equiv \dot{\rho}$ is the generator of the time evolution, while $D[\rho]$ denotes the dissipator, i.e., the non-unitary component responsible for the irreversible relaxation. In Lindblad form,

$$\mathcal{D}[\rho] = \sum_{\alpha,\beta} C_{\alpha\beta}(L_\alpha \rho L_\beta^\dagger - \frac{1}{2}\{L_\beta^\dagger L_\alpha, \rho\}), \tag{S7}$$

where $C_{\alpha\beta}$ is a positive semidefinite Kossakowski matrix. For spin systems, it is convenient to choose a Hermitian operator basis $F_\alpha = F_\alpha^\dagger$ for the dissipative term that includes system-bath couplings or spin noise; any non-Hermitian pieces can be incorporated into $H$ if necessary. In this basis, the dissipator can be rewritten in terms of Hermitian $F_\alpha$ and a real symmetric matrix $R_{\alpha\beta}$. To show this, we expand the commutator

$$[F_\beta, \rho] = F_\beta \rho - \rho F_\beta \tag{S8}$$

and then insert into a second commutator

$$\begin{aligned} \big[F_\alpha, [F_\beta, \rho]\big] &= F_\alpha(F_\beta\rho - \rho F_\beta) - (F_\beta\rho - \rho F_\beta)F_\alpha \\ &= F_\alpha F_\beta \rho - F_\alpha \rho F_\beta - F_\beta \rho F_\alpha + \rho F_\beta F_\alpha. \end{aligned} \tag{S9}$$

Multiplying by $-\frac{1}{2}R_{\alpha\beta}$ and summing over $\alpha, \beta$ gives

$$\mathcal{D}[\rho] = \sum_{\alpha,\beta} R_{\alpha\beta}(F_\alpha \rho F_\beta - \frac{1}{2}\{F_\alpha F_\beta, \rho\}) \tag{S10}$$

after a relabeling of dummy indices and using symmetry of $R_{\alpha\beta}$. This is equivalent to the standard Lindblad/Kossakowski form, with $L_\alpha = F_\alpha$ and $C_{\alpha\beta} = R_{\alpha\beta}$. The matrices $C_{\alpha\beta}$ and $R_{\alpha\beta}$ are related by a change of basis and carry the same physical information. Rearranging, one can show (see §2.2) that the dissipator can be written as

$$\mathcal{D}[\rho] = -\frac{1}{2}\sum_{\alpha,\beta} R_{\alpha\beta}[F_\alpha, [F_\beta, \rho]] \tag{S11}$$

with $R_{\alpha\beta}$ real, symmetric, and positive semidefinite, which is the Onsager (double commutator) form of the dissipator. Physically, $R_{\alpha\beta}$ is a relaxation tensor in the operator basis $\{F_\alpha\}$.

The matrix $R_{\alpha\beta}$ thus plays the role of a relaxation tensor in operator space, specifying how fluctuations of different operators contribute to dissipation. At the level of the density operator $\rho$, this distinction corresponds to a decomposition of the generator of motion into an antisymmetric part, $[H, \rho]$, describing unitary evolution, and a symmetric part, proportional to $R_{\alpha\beta}$, describing irreversible relaxation and entropy production. The most general Markovian evolution may therefore be written schematically as

$$\dot{\rho} = -\frac{i}{\hbar}[H, \rho] - \frac{1}{2}\sum_{\alpha,\beta} R_{\alpha\beta}([F_\alpha, [F_\beta, \rho]]) \tag{S12}$$

where the commutator term encodes parallel transport under a connection contained in the Hamiltonian, and the double commutator encodes dissipation and entropy production. In the geometric picture developed below, we will see that the commutator term corresponds to parallel transport along the Hamiltonian flow, while the symmetric part defines a dissipative metric structure.

## 2.2 The commutator of covariant derivatives

The dissipator developed in §2.1 contracts successive adjoint actions with the symmetric tensor $R_{\alpha\beta}$, so their antisymmetric part cancels. The antisymmetric difference between the two possible orders of covariant transport gives the curvature operator $\left[\nabla_\mu, \nabla_\nu\right]\rho$. We define the covariant derivative on the density operator as

$$\nabla_\mu \rho \equiv \partial_\mu \rho + \frac{i}{\hbar}[A_\mu, \rho], \tag{S13}$$

where $A_\mu$ is a connection and the index $\mu$ labels a direction in the parameter space over which it is defined. Familiar instances include the electromagnetic, Lorentz, and Levi–Civita connections over spacetime, and the Berry connection over a manifold of control parameters. The connection developed here is of the latter type, but acts in Liouville space on the full generator rather than accumulating a phase on states under adiabatic unitary evolution. We wish to show

$$\left[\nabla_\mu, \nabla_\nu\right]\rho = \frac{i}{\hbar}\left[F_{\mu\nu}, \rho\right] \tag{S14}$$

with

$$F_{\mu\nu} = \partial_\mu A_\nu - \partial_\nu A_\mu + \frac{i}{\hbar}[A_\mu, A_\nu]. \tag{S15}$$

Begin by writing

$$\nabla_\mu = \partial_\mu + \frac{i}{\hbar}\, ad_{A_\mu}, ad_{A_\mu}(\cdot) \equiv [A_\mu, \cdot]. \tag{S16}$$

Acting on $\rho$ gives

$$\nabla_\nu \rho = \partial_\nu \rho + \frac{i}{\hbar}[A_\nu, \rho]. \tag{S17}$$

Next, computing $\nabla_\mu(\nabla_\nu \rho)$ yields

$$\begin{aligned} \nabla_\mu(\nabla_\nu \rho) \quad &= \partial_\mu(\partial_\nu \rho + \tfrac{i}{\hbar}[A_\nu, \rho]) + \tfrac{i}{\hbar}[A_\mu, \partial_\nu \rho + \tfrac{i}{\hbar}[A_\nu, \rho]] \\ &= \partial_\mu \partial_\nu \rho + \tfrac{i}{\hbar}\partial_\mu [A_\nu, \rho] + \tfrac{i}{\hbar}[A_\mu, \partial_\nu \rho] - \tfrac{1}{\hbar^2}[A_\mu, [A_\nu, \rho]]. \end{aligned} \tag{S18}$$

Similarly,

$$\nabla_\nu(\nabla_\mu \rho) \quad = \partial_\nu \partial_\mu \rho + \tfrac{i}{\hbar}\partial_\nu [A_\mu, \rho] + \tfrac{i}{\hbar}[A_\nu, \partial_\mu \rho] - \tfrac{1}{\hbar^2}[A_\nu, [A_\mu, \rho]]. \tag{S19}$$

Subtracting gives the commutator:

$$[\nabla_\mu, \nabla_\nu]\rho = \nabla_\mu(\nabla_\nu \rho) - \nabla_\nu(\nabla_\mu \rho). \tag{S20}$$

The $\partial_\mu \partial_\nu \rho$ and $\partial_\nu \partial_\mu \rho$ terms cancel because partial derivatives commute. Grouping the remaining pieces, we gather:

1. Derivative terms:

$$\frac{i}{\hbar}(\partial_\mu [A_\nu, \rho] - \partial_\nu [A_\mu, \rho]) = \frac{i}{\hbar}([\partial_\mu A_\nu - \partial_\nu A_\mu, \rho]) + \frac{i}{\hbar}([A_\nu, \partial_\mu \rho] - [A_\mu, \partial_\nu \rho]). \tag{S21}$$

2. The $[A, \partial\rho]$ terms cancel against those coming from $[A_\mu, \partial_\nu \rho]$ and $[A_\nu, \partial_\mu \rho]$ in the second line.

3. The double-commutator terms combine using the Jacobi identity:

$$-\frac{1}{\hbar^2}\left([A_\mu,[A_\nu,\rho]]-[A_\nu,[A_\mu,\rho]]\right)=-\frac{1}{\hbar^2}[[A_\mu,A_\nu],\rho]. \quad \text{(S22)}$$

Collecting everything,

$$[\nabla_\mu,\nabla_\nu]\rho=\frac{i}{\hbar}\left[\partial_\mu A_\nu-\partial_\nu A_\mu+\frac{i}{\hbar}[A_\mu,A_\nu],\,\rho\right]=\frac{i}{\hbar}[F_{\mu\nu},\rho]. \quad \text{(S23)}$$

This immediately echoes the geometrodynamics statement that curvature is a commutator of covariant derivatives, here written in the adjoint representation acting on $\rho$.[5]

## 2.3 The rate matrix $R_{ij}$: Bloch picture

Expectation values of observables $S_i = \mathrm{Tr}(\rho\sigma_i)$ obey a closed equation of motion obtained by projecting the master equation onto the Pauli basis. Under the same weak coupling and Markovian assumptions, the microscopic superoperator defined by $R_{\alpha\beta}$ produces linear relaxation of the Bloch vector,

$$\dot{S}_i=(\Omega\times S)_i-\sum_j R_{ij}(S_j-S_{j,eq}), \quad \text{(S24)}$$

where $R_{ij}$ is the representation of $R_{\alpha\beta}$ in the Pauli (adjoint SU(2)) basis. In the conventional Bloch form (laboratory frame, secular approximation), relaxation acts as a linear map on the magnetization vector,

$$\dot{M}=M\times\omega_0-\begin{pmatrix}1/T_2&0&0\\0&1/T_2&0\\0&0&1/T_1\end{pmatrix}(M-M_{\mathrm{eq}}), \quad \text{(S25)}$$

with $M_{\mathrm{eq}}=(0,0,M_0)$. Thus, the standard Bloch rates correspond simply to the diagonal form $R_{ij}=diag(R_2,R_2,R_1)$. In this representation the relaxation operator already appears as a tensor acting on the state vector, although expressed in a particular coordinate frame.

## 2.4 A covariant formulation

To express relaxation in a coordinate-independent form, we embed the three-component Bloch vector into a four-component state vector $S^\mu$ defined in a chosen frame. The temporal component $S^0$ represents the conserved scalar amplitude, while the spatial components $S^{i(i=1,2,3)}$ correspond to transverse coherence and longitudinal population, respectively. In this representation, the relaxation matrix $R_{ij}$ is naturally extended to a rank-2 tensor acting on the state space,

$$R^\mu{}_\nu=\begin{pmatrix}0&0&0&0\\0&1/T_2&0&0\\0&0&1/T_2&0\\0&0&0&1/T_1\end{pmatrix} \quad \text{(S26)}$$

which makes explicit that longitudinal and transverse relaxation arise as different projections of the same operator.

The evolution law may then be written compactly as

$$\dot{S}^\mu=\Omega^\mu{}_\nu S^\nu-R^\mu{}_\nu\left(S^\nu-S^\nu_{eq}\right) \quad \text{(S27)}$$

where $\Omega^\mu{}_\nu$ generates unitary precession (the phase contributions from electromagnetic field, Lorentz, Levi-Civita, or other control-driven connections) and $R^\mu{}_\nu$ encodes dissipation. In the laboratory frame and Bloch limit, this reduces directly to the familiar rates $R_1$ and $R_2$ and their standard action on $\mathbf{M}$. Yet, the tensor form is useful because it also accommodates non-diagonal relaxation when the bath correlation tensor is anisotropic or when non-secular terms mix longitudinal and transverse components. We emphasize that "covariant" here does not imply Lorentz invariance in the usual NMR sense; rather, $R^{\mu}{}_{\nu}$ is a single tensorial object whose projections recover the observed longitudinal and transverse decay, as well as any experimentally relevant mixing between them.

Spin precession can be treated as parallel transport under a connection $A_\mu$ that produces holonomy. On density matrices, parallel transport is written as a covariant derivative acting by commutator:

$$\nabla_\mu \rho \equiv \partial_\mu \rho + \frac{i}{\hbar}\left[A_\mu, \rho\right]. \tag{S28}$$

One can then show (see § 2.2) that

$$\left[\nabla_\mu, \nabla_\nu\right]\rho = \frac{i}{\hbar}\left[F_{\mu\nu}, \rho\right], \tag{S29}$$

for a curvature 2-form $F_{\mu\nu}$, and

$$F_{\mu\nu} = \partial_\mu A_\nu - \partial_\nu A_\mu + \frac{i}{\hbar}[A_\mu, A_\nu]. \tag{S30}$$

This identity (shown pictorially in Fig. 1 of the main text) links commutators to curvature in the language of transport on the bundle of spin states (or on the operator algebra), in a way that is well-established in gauge theory. Combined with dissipation, the full time evolution is

$$\nabla_t \rho = \mathcal{D}[\rho], \tag{S31}$$

where $\nabla_t$ is the covariant time derivative (including geometric/holonomic transport) and $\mathcal{D}[\rho]$ is the Lindbladian. This isolates the different roles plainly – curvature is contained in the operator $\left[\nabla_\mu, \nabla_\nu\right]$, governing reversible holonomy, and dissipation contains a double commutator that mediates the irreversible loss of distinguishability (entropy production) under the same covariant derivatives that define curvature.

## 2.5 From Lindblad dissipation to a relaxation tensor $R_{\mu\nu}$

As introduced in §2.1, a Markovian, positive, trace-preserving generator can always be written in Lindblad or Gorini-Kossakowski-Sudarshan-Lindblad (GKSL) form as

$$\mathcal{D}(\rho) = \sum_k \left(L_k \rho L_k^\dagger - \frac{1}{2}\left\{L_k^\dagger L_k, \rho\right\}\right), \tag{S32}$$

where $\{L_k\}$ are Lindblad operators and $\{\cdot,\cdot\}$ denotes the anticommutator. A fundamental claim of the covariant relaxation tensor viewpoint is that the irreversible part of this generator is represented as a symmetric positive semidefinite bilinear form on Liouville space (operator space), which then becomes the object we call $R$.

To make this specific, we begin by fixing an operator basis $\{F_a\}$ for the relevant operator space (e.g., traceless operators plus identity, or the full matrix algebra). Choosing it to be orthonormal under the Hilbert–Schmidt inner product gives

$$\langle A, B\rangle := \mathrm{Tr}\left(A^\dagger B\right), \tag{S33}$$

with

$$\mathrm{Tr}(F_\alpha^\dagger F_\beta) = \delta_{\alpha\beta}. \tag{S34}$$

Expanding each Lindblad operator in this basis yields

$$L_k = \sum_\alpha l_{k\alpha}\, F_\alpha. \tag{S35}$$

with coefficient vectors $l_k$ and summation index $\alpha$. The Kossakowski matrix can be defined as

$$C_{\alpha\beta} := \sum_k l_{k\alpha} l_{k\beta}^*, \tag{S36}$$

which is Hermitian and positive semidefinite by construction, because it is a Gram matrix of coefficient vectors $l_k$. This means that for any complex vector $x_\alpha$,

$$x_\alpha^*\, C_{\alpha\beta}\, x_\beta \geq 0. \tag{S37}$$

Using this basis expansion, the GKSL dissipator can be written as

$$\mathcal{D}(\rho) = \sum_{\alpha\beta} C_{\alpha\beta} \left( F_\alpha \rho F_\beta^\dagger - \frac{1}{2} \left\{ F_\beta^\dagger F_\alpha, \rho \right\} \right). \tag{S38}$$

At this point, one introduces a standard decomposition of $C$ into symmetric and antisymmetric parts. The form of that decomposition depends slightly on whether one chooses Hermitian basis operators $F_\alpha = F_\alpha^\dagger$, as is common in spin systems, or a general complex basis. Conceptually, the part of the quadratic generator that is antisymmetric in the operator indices contributes to an effective Hamiltonian correction, analogous to a Lamb-shift commutator term. Meanwhile, the symmetric part contributes to a double-commutator contraction term.

In the case of a Hermitian basis, which is aligned with the double commutator framing, one may write

$$C_{\alpha\beta} = C_{\alpha\beta}^{(S)} + C_{\alpha\beta}^{(A)}, \tag{S39}$$

$$C_{\alpha\beta}^{(S)} = C_{\beta\alpha}^{(S)}, \tag{S40}$$

$$C_{\alpha\beta}^{(A)} = -C_{\beta\alpha}^{(A)}. \tag{S41}$$

where the (S) and (A) superscripts indicate symmetric and antisymmetric parts, respectively. The dissipator can then be rearranged into

$$\mathcal{D}(\rho) = -\frac{1}{2} \sum_{\alpha\beta} C_{\alpha\beta}^{(S)} \, \left[ F_\alpha, \left[ F_\beta, \rho \right] \right], \tag{S42}$$

and applied in conjunction with some Hermitian $H$ constructed from the antisymmetric part $C^{(A)}$. The specific expression for $H$ is not important for the present purpose; what matters is that it is Hamiltonian-like and therefore belongs to the antisymmetric, i.e. the connection, part of the generator to construct the full Liouvillian.

Now relaxation tensor object can be defined canonically as the symmetric bilinear form

$$R_{\alpha\beta} := C_{\alpha\beta}^{(S)}. \tag{S43}$$

Thus, as similarly written in § 2.1,

$$\mathcal{D}(\rho) = -\frac{1}{2} \sum_{\alpha\beta} R_{\alpha\beta} \left[F_\alpha, \left[F_\beta, \rho\right]\right]. \tag{S44}$$

The core structural properties of $R_{\alpha\beta}$ are inherited immediately. First, it is symmetric:

$$R_{\alpha\beta} = R_{\beta\alpha}, \tag{S45}$$

and second, it is positive semidefinite:

$$x_\alpha^* \, R_{\alpha\beta} \, x_\beta \geq 0. \tag{S46}$$

The positive semidefinite property is what makes this object "relaxation-like" in the strongest sense. It is a contraction form on operator space, and it is the precise algebraic avatar of complete positivity in the symmetric sector. At the semantic level, this is the intended meaning of the statement that "the dissipator corresponds to a covariant relaxation tensor": once an operator frame is chosen, the GKSL dissipator induces a symmetric positive semidefinite bilinear form $R_{\alpha\beta}$ such that the irreversible generator can be written as a contraction (double commutator) weighted by $R$, while the antisymmetric residue is reabsorbed into a Hamiltonian/connection term.

## 2.6 Index notation and coordinate frames

Here we employ two kinds of indices: the operator frame indices $a, b$ (or $\alpha, \beta$) as well as the physical, covariant indices $\mu, \nu$. The notation $R_{\alpha\beta}$ above should be understood as the representation of the relaxation form in a chosen operator basis $\{F_a\}$. These indices $a, b$ are thus "internal" Liouville indices because they label directions in operator space.

Separately, when we write $R_{\mu\nu}$ in the covariant narrative (e.g., when connecting dynamics to continuity law forms, or when embedding into broader field equation analogies), the indices $\mu, \nu$ label components in some physically organized coordinate chart for the reduced state variables. Examples include (i) Bloch coordinates $S_i = \mathrm{Tr}(\rho\,\sigma_i)$, where $\mu = i \in \{x, y, z\}$; (ii) extended coordinate sets that include energy-like and entropy-like components in addition to spin polarization components, and (iii) any other vector of slow variables regarded as the relevant coarse-grained state.

The salient point here is that $R_{\alpha\beta}$ and $R_{\mu\nu}$ do not represent different physics – they are the same bilinear form expressed in different coordinate systems. To state this cleanly, we can introduce a linear map between the operator frame components and the physical coordinate components. Abstractly, there exists a change-of-basis matrix ${E_\mu}^a$ such that the covariant components are obtained by the transformation

$$R_{\mu\nu} = {E_\mu}^a \, R_{\alpha\beta} \, {E_\nu}^b. \tag{S47}$$

Here the role of ${E_\mu}^a$ is to "push" $R_{\alpha\beta}$ from the operator frame to a different coordinate basis; the $\mu$-coordinate direction corresponds to a particular linear combination of operator basis directions.

Generally, this is the same transformation law that any (0, 2) tensor obeys under a change of basis. This is also why it is reasonably consistent to call the object "covariant," because the physical information it contains is independent of the basis, while the component matrix changes by the appropriate tensor transformation.

It is worth emphasizing what is (and what is not) being posited. The positive semidefinite property originates in the operator space representation $R_{\alpha\beta}$ because that is where complete positivity lives most directly, via the Kossakowski matrix. When one pushes forward into a reduced physical coordinate set, the inherited positivity statement becomes a statement about the induced quadratic form on that reduced set, provided the reduction respects the relevant subspace. In practice, if $\delta X^\mu$ is a vector of small deviations in the chosen physical coordinates, the contraction $\delta X^\mu R_{\mu\nu} \delta X^\nu$ is the inherited nonnegative scalar quantity.

### 2.7 $R_{\mu\nu}$ as a $(0,2)$ tensor versus $R^\mu{}_\nu$ as a $(1,1)$ tensor

There are two related but unique roles that $R$ plays. First, $R$ acts as a bilinear form defining contraction and, in off-diagonal components, exchange between entropy production and energy. Second, $R$ is as a linear map acting on a deviation vector in the equation of motion. These roles correspond to different index placements.

Treating $R$ as a $(0,2)$ tensor means treating it as something that takes two vectors and returns a scalar (i.e., a bilinear form). This is the representation that aligns immediately with complete positivity via the symmetric Kossakowski form, Onsager-type dissipation functionals, entropy production as a nonnegative scalar, and the ability to interact with other $(0,2)$ tensors in field equations, since most geometric objects in those settings are written with lower indices and contracted into scalars. In this view, $R_{\mu\nu}$ is the fundamental object.

Although it is natural to treat $R$ as a $(0,2)$ tensor $R_{\mu\nu}$, a $(1,1)$ version $R^\mu{}_\nu$ (i.e., an endomorphism) is also necessary for dynamics and commutators. The relaxation term in a first order equation of motion is naturally written as an object that acts on a vector of deviations, schematically

$$(\text{relaxation term})^\mu = -R^\mu{}_\nu \, \delta X^\nu. \tag{S48}$$

To obtain $R^\mu{}_\nu$ from $R_{\mu\nu}$, one must choose a metric $g_{\mu\nu}$, or more generally, an inner product identification between covectors and vectors, on the relevant state coordinate space and raise an index:

$$R^\mu{}_\nu := g^{\mu\alpha} R_{\alpha\nu}. \tag{S49}$$

This is necessary if one wishes to write commutators of generators in a clean way, because commutators are naturally defined on endomorphisms. To write the commutator $[K, R]$ for conserved vector field $K$ (i.e., a Killing vector, developed in § 4), the necessary mathematical object is

$$\big[K, R\big]^\mu{}_\nu := K^\mu{}_\alpha R^\alpha{}_\nu - R^\mu{}_\alpha K^\alpha{}_\nu \,, \tag{S50}$$

which requires $R$ to be viewed as a $(1,1)$ linear map. The underlying bilinear form of the covariant relaxation tensor is $R_{\mu\nu}$; here, the $(1,1)$ representation serves as the acting version obtained after choosing the metric.

A reasonable convention is thus to define and interpret $R_{\mu\nu}$ as the covariant relaxation tensor, a bilinear form, and then to introduce $R^\mu{}_\nu$ via a chosen metric $g^{\mu\nu}$ whenever one needs an operator to act on state deviations or to evaluate commutators. This is a concise way to maintain both conceptual clarity and index consistency.

### 2.8 Connecting $R^\mu{}_\nu$ to Redfield theory: projections onto driven trajectories

In the Redfield limit, the components of the relaxation operator are determined microscopically by bath correlation functions and their spectral densities. Writing the system–bath coupling as

$$H_1(t) = -\gamma \mathbf{S} \cdot \delta \mathbf{B}(t), \tag{S51}$$

with

$$\delta\mathbf{B}(t) = \left(\delta B_x(t), \delta B_y(t), \delta B_z(t)\right), \tag{S52}$$

Redfield theory yields relaxation rates built from the spectral densities $J_{ij}(\omega)$ of the fluctuating field components. In the conventional (weak-coupling, secular) Bloch regime, the familiar scalars $R_1$ and $R_2$ are recovered as projected components of the relaxation operator:

$$R_{zz} = R_1 = \gamma^2 \left[J_{xx}(\omega_0) + J_{yy}(\omega_0)\right] \tag{S53}$$

$$R_{xx} = R_{yy} = R_2 = \frac{1}{2}R_1 + \gamma^2 J_{zz}(0). \tag{S54}$$

Equivalently, one may regard Redfield theory as providing the constitutive law for the elements of $R_{ij}$ (and, in the more general case, for $R^{\mu}{}_{\nu}$, in the same sense that conductivity or permittivity supplement Maxwell's equations – the field equations define the structure of the dynamics, while the relaxation tensor is material- and environment-dependent).

This interpretation reveals why the diagonal Bloch matrix is so often adequate in conventional practice. When the noise is effectively isotropic and the secular approximation holds, the relaxation tensor is diagonal in the same frame used to observe the dynamics. When these conditions fail – in anisotropic baths, tilted effective fields, or non-secular mixing – the full tensor structure becomes relevant and can be observed experimentally.

When the system is driven along a lower-dimensional manifold of control parameters $\lambda^i$ (pulse phases, flip angles, effective Bloch sphere coordinates), the tensor induces an effective quadratic form on that control space,

$$\Gamma_{ij}(\lambda) = \frac{\partial x^{\mu}}{\partial \lambda^i}\frac{\partial x^{\nu}}{\partial \lambda^j} R_{\mu\nu}(x(\lambda)). \tag{S55}$$

Along a trajectory $\lambda^i$, the instantaneous dephasing rate becomes

$$\frac{1}{T_2^{eff}(\lambda, \dot{\lambda})} = \Gamma_{ij}(\lambda)\, \dot{\lambda}^i \dot{\lambda}^j, \tag{S56}$$

showing that decoherence is the contraction of the relaxation tensor with the velocity of the state through control space. In isotropic cases, this reduces to a constant $T_2$. For driven experiments with nontrivial paths, the rate becomes path- and speed-dependent.

Contractions of this kind underlie the angular dependence of conventional relaxation measurements. Two familiar cases are $R_{1\rho}$, where the relevant angle is the tilt $\theta$ of the effective field, and $R_2$ anisotropy in oriented solids, where the angle is that between $B_0$ and the local symmetry axis of the coupling. In each, a symmetric second-rank tensor is contracted along a measurement direction and the measured rate varies with the angle between them. Such a dependence on a single angle can be expanded in Legendre polynomials $P_L$, and only even orders $L$ appear here because both couplings are invariant under inversion.[6] The structure of the measurement determines which orders $L$ arise in a given experiment. The two cases here differ because the contraction and the second-order step of Redfield theory occur in opposite orders. Relaxation rates are second order in the system–bath coupling since they are built from the spectral densities $J_{ij}(\omega)$, which are Fourier transforms of two-point correlation functions of the fluctuating field.

For $R_{1\rho}$, the relaxation tensor is already a second-order object, and it is contracted once along the effective-field axis. With $\cos^2\theta = \frac{1}{3}(1 + 2P_2)$ and $\sin^2\theta = \frac{2}{3}(1 - P_2)$,

$$R_1 \cos^2\theta + R_2 \sin^2\theta = \frac{R_1 + 2R_2}{3}\, P_0 + \frac{2}{3}\,(R_1 - R_2)\, P_2(\cos\theta), \tag{S57}$$

so the tilt dependence contains $L = 0$ and $L = 2$ only.

For orientation-dependent $R_2$, the contraction is applied earlier, to the dipolar coupling tensor onto the measurement frame. For two spins separated by an internuclear vector $\mathbf{r} = r\hat{u}$, the dipolar coupling tensor is $D_{ij} = b\left(\delta_{ij} - 3\hat{u}_i\hat{u}_j\right)$ with $b = \mu_0\gamma_I\gamma_S\hbar/4\pi r^3$.[7] Secular truncation with respect to $B_0$ retains only the component of $D$ along the field,

$$\omega_D = D_{zz} = \widehat{B}_0^{\intercal} \cdot D \cdot \widehat{B}_0 = -2b\, P_2\left(\cos\beta\right), \tag{S58}$$

where $\beta$ is the angle between $\hat{u}$ and $B_0$. Where molecular motion is fast and axially symmetric about a local director, this form is retained with $\beta$ the angle between the director and $B_0$, and $b$ scaled by an order parameter.[8] The rate is then second order in the projected coupling, $R_2 \propto \langle\delta\omega_D^2\rangle\,\tau_c$, with the angular factor squared. With $x = \cos\beta$, $P_2(x)^2 = \frac{1}{4}\left(9x^4 - 6x^2 + 1\right)$, and substituting $x^2 = \frac{1}{3}P_0 + \frac{2}{3}P_2$ and $x^4 = \frac{1}{5}P_0 + \frac{4}{7}P_2 + \frac{8}{35}P_4$,

$$\left[P_2\left(\cos\beta\right)\right]^2 = \frac{1}{5}P_0 + \frac{2}{7}P_2 + \frac{18}{35}P_4, \tag{S59}$$

equivalently the squared Clebsch–Gordan coefficients $\left(C^{L0}_{20,20}\right)^2$ for $L = 0, 2, 4$, which sum to unity as required by $P_2(1) = 1$.[6]

An $L = 4$ term therefore appears when the contraction precedes the second-order step. This holds when transverse decay is relaxation-dominated and the motional averaging is axially symmetric. In ordered systems, the measured decay typically also contains a coherent contribution from the residual static coupling, which can produce a fourth-order term of its own. Fitting a decay rate to a distribution of static couplings returns a second-moment quantity, and squaring $\omega_D \propto P_2\left(\cos\beta\right)$ reintroduces $P_4$. Hence, measuring the decay while rotating the sample about a single axis cannot attribute a $P_4$ term, because both mechanisms can produce one. Since the rotation pattern alone is underdetermined, the standard tools separate the relaxation and coherent decay contributions by other means. The coherent part refocuses under an echo and averages under magic angle spinning, and varying echo spacing $\tau$ reveals whether the $P_4$ amplitude tracks the relaxation rate or the coupling.

## 2.9 Geometric interpretation and consequences: transport, currents, and entropy production

The tensor formulation above suggests a geometric picture of relaxation. Rather than treating $R_1$ and $R_2$ as independent decay constants, they might be viewed as different projections of a single dissipative motion of the spin state on the Bloch sphere. In this view, relaxation is not the action of separate mechanisms but the gradual transport of information and energy across the state manifold.

To make this notion concrete, we employ a scalar phase potential $\Phi$ whose gradients generate an information (entropy) current

$$J_S^{\mu} \equiv \beta\nabla^{\mu}\Phi, \tag{S60}$$

where, $\beta \sim \pi/k_B$ sets the scale connecting phase dispersion to entropy flow. Dissipative effects are then controlled by the divergence of this current,

$$\nabla_{\mu}J_S^{\mu} = \chi\Theta\beta\Box\Phi \tag{S61}$$

where $\Theta$ represents a bath temperature scale and $\chi$ is a coupling strength that sets the overall relaxation rate. In this form, dissipation appears as diffusion on the Bloch sphere manifold, curvature of the phase landscape drives irreversible spreading, and the temperature scale controls the speed of that spreading.

## 2.10 Spin echoes as a separation of reversible and irreversible transport

Spin echoes provide an illustration of these phenomena. Components whose accumulated phase cancels over an echo cycle retrace their paths and rephase, isolating a reversible part of the dynamics associated with static inhomogeneity or deterministic evolution that is, in principle, refocusable. The remaining decay that cannot be recovered by unitary pulse control marks true entropy production, i.e., information that has flowed irreversibly into environmental degrees of freedom. In the transport language, refocusable parts contribute to the divergence-free component, while irreversible decay corresponds to a nonzero divergence of an information current.

## 2.11 Parallel transport and a geometric view of commutators

This geometric intuition clarifies the meaning of commutators. In classical differential geometry, the commutator of two vector fields measures the failure of successive transports to commute: flowing along one field and then another does not, in general, return to the same point as flowing along the second before the first. This "failure to close" is a kinematic signature of curvature.

Figure 1 in the main text is a pictorial representation of the commutator as a closer of quadrilaterals, for generalized vector fields as well as an adapted view for Liouville-space vector fields (superoperators). Fig. 1a shows two vector fields $\mathbf{u}(\mathscr{P})$ and $\mathbf{v}(\mathscr{P})$ in spacetime; at the event $\mathscr{P}_0$, the commutator $[\mathbf{u},\mathbf{v}]$ is denoted by the vector $\mathscr{P}_4 - \mathscr{P}_3$, which measures the degree to which the four-sided surface between events $\mathscr{P}_1$, $\mathscr{P}_2$, $\mathscr{P}_3$, and $\mathscr{P}_4$ fails to close. In a coordinate basis,

$$\mathscr{P}_4 - \mathscr{P}_3 = [\mathbf{u}(\mathscr{P}_0) + \mathbf{v}(\mathscr{P}_1)] - [\mathbf{u}(\mathscr{P}_2) + \mathbf{v}(\mathscr{P}_0)] \tag{S62}$$

$$= [\mathbf{v}(\mathscr{P}_1) - \mathbf{v}(\mathscr{P}_0)] - [\mathbf{u}(\mathscr{P}_2) - \mathbf{u}(\mathscr{P}_0)] \tag{S63}$$

$$= \left(v^{\beta}{}_{,\alpha} u^{\alpha} \mathbf{e}_{\beta}\right)_{\mathscr{P}_0} - \left(u^{\beta}{}_{,\alpha} v^{\alpha} \mathbf{e}_{\beta}\right)_{\mathscr{P}_0} + errors \tag{S64}$$

$$= [\mathbf{u},\mathbf{v}]_{\mathscr{P}_0} + errors \tag{S65}$$

where errors are terms such as $v^{\beta}{}_{,\mu\nu} u^{\mu} u^{\nu} \mathbf{e}_{\beta}$ (c.f. Ref. [5], Box. 9.2). In this kinematics picture, we compare the order-dependence of flow along $\mathbf{u}$, then flow along $\mathbf{v}$. In quantum mechanics, superoperators $\mathscr{L}$ act as vector fields in Liouville space, with events $\mathscr{P}$ replaced by the density matrix $\rho$ in Fig. 1b. Here, the "closer of the quadrilateral" is the double commutator $[[\mathscr{H}_1, \mathscr{H}_2], \rho_0]$ for $\mathscr{L}_1(\rho_0) = -\frac{i}{\hbar}[\mathscr{H}_1, \rho_0]$, $\mathscr{L}_2(\rho_0) = -\frac{i}{\hbar}[\mathscr{H}_2, \rho_0]$, and hence $[\mathscr{L}_1, \mathscr{L}_2](\rho_0) = -\frac{1}{\hbar^2}[[\mathscr{H}_1, \mathscr{H}_2], \rho_0]$. The effect of relaxation is to shrink the quadrilateral irreversibly. The commutator of two Liouville-space generators is again the closure defect, now expressed as a double commutator on $\rho$. In this way, the geometry of transport connects algebraic commutators to the curvature of the control connection, developed further in §§3.2–3.3 of this SI.

## 2.12 Physical meaning of off-diagonal relaxation

The diagonal elements of $R$ describe decay of each component along itself. Off-diagonal elements describe how dissipation mixes components. They represent direction-dependent couplings imposed by the environment rather than additional loss pathways. Terms such as $R_{xz}$ or $R_{yz}$ couple longitudinal and transverse magnetization, and $R_{xy}$ mixes in-phase and quadrature transverse components. Such terms arise when the dissipative operators are anisotropic, or when the basis that diagonalizes dissipation does not coincide with the basis in which the system is driven and observed (illustrated in Fig. 2 in the main text). Conventional NMR is often designed to operate in regimes where these terms are averaged away – rapid isotropic tumbling, motional narrowing, magic-angle spinning, or alignment of dressed and laboratory axes under on-resonance driving – which is one reason the tensor viewpoint is not usually emphasized.

When symmetry is reduced – in anisotropic solids, surface-bound spins, paramagnetic systems, tilted effective fields under detuning or spin-lock conditions, or pulse trains that explore multiple non-commuting directions – the off-diagonal structure becomes measurable and can dominate signal envelopes. In these settings, the tensor structure provides an adept language for describing relaxation as a geometry-dependent flow on the state space.

Importantly, the off-diagonal structure is a property of the material and its environment, not of the pulse sequence. Pulse sequences cannot create such terms; they can only average them away or reveal them. The tensor $R_{\mu\nu}$ is therefore intrinsic, while the components that are measured depend on how the system is probed.

There has long been a tension between the seemingly irreconcilable juxtaposition that quantum mechanics is strictly unitary while thermodynamics is irreversibly entropic. The intermediate regime involving decoherence and information loss is perhaps the most confounding of them all. Yet, a different view of spin relaxation suggests a structural bridge between these domains. The interpretation developed here recasts decoherence as geometric entropy flow, with $R_1$ and $R_2$ arising from different projections of one jointly conserved tensorial object. This interpretation may prove useful in the future if it not only links holonomy at the level of a single spinor and phase scrambling in a statistical ensemble, but is also operative in thermodynamic systems at macroscopic scales.

# 3 Symmetry development with Killing vector formalism

## 3.1 Formal development of conserved generators in spin dynamics via Killing vectors

In general relativity, a Killing vector field (so named after Wilhelm Killing) generates a symmetry of the metric or, more broadly, a symmetry of the system.[5] In spectroscopy, Killing vectors appear mathematically as continuous symmetries that leave the Hamiltonian invariant. In quantum mechanics, the analog is

$$\mathcal{L}_K \mathcal{H} = 0 \Longleftrightarrow [K, \mathcal{H}] = 0 \tag{S66}$$

where $\mathcal{L}_K$ is the Liouvillian superoperator, $\mathcal{H}$ is the Hamiltonian, and $K$ the corresponding Killing vector that is conserved under parallel transport. To illustrate the formalism, below we consider some relevant cases for NMR, where the relevant Killing vectors are identified as the generators of the symmetries of the spin Hamiltonian.

### 3.1.1 Static Zeeman Hamiltonian

For a spin in a strong static field $B_0\hat{z}$, the Zeeman Hamiltonian is

$$H_Z = -\gamma B_0 I_z. \tag{S67}$$

The symmetry here is that the Hamiltonian is invariant under rotations about the z-axis. The generator for the Killing vector is

$$K = I_z \tag{S68}$$

because $[I_z, H_Z] = 0$. Thus, the Killing vector corresponds to $\partial_\phi$, which is a rotation around z (using the definition of vectors as directional derivatives). This is axial $U(1)$ symmetry. Physically, this means that magnetization along z is conserved, in absence of relaxation.

### 3.1.2 Isotropic J-coupling

For two J-coupled spins, $\mathbf{I}_1$ and $\mathbf{I}_2$ with coupling constant $J$, the Hamiltonian

$$H_J = 2\pi J \mathbf{I}_1 \cdot \mathbf{I}_2 \tag{S69}$$

is rotationally invariant. Here there is full 3D rotation symmetry. The generators for the Killing vectors are

$$K_i = I_{1i} + I_{2i}, \tag{S70}$$

with $i = x, y, z$. They satisfy $[K_i, H_J] = 0$, and the symmetry group is $SO(3)$. Physically, this means that total angular momentum is conserved.

### 3.1.3 Secular Dipolar Hamiltonian (High Field)

Under high field truncation, the dipolar Hamiltonian is proportional to

$$H_D \propto 3I_{1z}I_{2z} - \mathbf{I}_1 \cdot \mathbf{I}_2. \tag{S71}$$

Here, rotational symmetry is broken by $B_0$. The only surviving symmetry is rotations about $z$. The generator for the Killing vector is

$$K = I_{1z} + I_{2z}, \tag{S72}$$

because only the z-components for each spin operator commute with the Hamiltonian. As with the Zeeman interaction, the Killing vector corresponds to $\partial_\phi$ and represents axial $U(1)$ symmetry. Physically, this means that in the presence of dipolar coupling, the external field breaks SO(3) symmetry.

### 3.1.4 Rf-Driven Hamiltonian

During rf pulse control, the time-dependent Hamiltonian can be written

$$H = -\gamma B_0 I_z - \gamma B_1 \left( I_x \cos \omega t + I_y \sin \omega t \right) \tag{S73}$$

where the only static interaction included here is the Zeeman term. In the rotating frame on resonance, this can be written more simply as

$$H_{\mathrm{rot}} = \Delta I_z + \omega_1 I_x. \tag{S74}$$

Here, symmetry depends on detuning: on resonance ($\Delta = 0$), there is symmetry about the x-axis. Off resonance, however, there is no continuous symmetry. Thus the Killing vector changes with frame and drive.

### 3.1.5 Gradient Hamiltonian

In the presence of a magnetic field gradient, there is an added term to the Hamiltonian

$$H = \gamma G_Z(t) z I_z \tag{S75}$$

where $G_Z(t)$ indicates the time-dependent gradient strength along z (although this approach is relevant for gradients along any or multiple dimensions). Here, this breaks translational symmetry in z; as a result, there is no spatial Killing vector along z. However, spin rotation symmetry about z remains, expressed via the generator

$$K = I_z. \tag{S76}$$

This means the internal spin symmetry survives while spatial symmetry is broken, and again, the Killing vector corresponds to $\partial_\phi$ and represents axial $U(1)$ symmetry. Physically, this means that the gradient acts like a background field breaking spatial isometry.

#### 3.1.6 Relaxation (Open System)

To include relaxation in this picture, it is useful to employ Redfield and Lindblad formalisms, and to work with spin densities in superoperator space:

$$\dot{\rho} = -i[H, \rho] + \mathcal{D}(\rho). \tag{S77}$$

The Liouville Killing generator for z-rotations can be written as

$$\mathcal{K}_z(\rho) = -i\,[I_z, \rho]\,. \tag{S78}$$

Symmetry requires that $[\mathcal{K}_z, \mathcal{L}] = 0$, which expands to:

$$[\mathcal{K}_z, \mathcal{L}_H] + [\mathcal{K}_z, \mathcal{D}] = 0. \tag{S79}$$

For a Hamiltonian with Zeeman and secular dipolar terms,

$$[\mathcal{K}_z, \mathcal{L}_H] = 0. \tag{S80}$$

The main question, then, is: $[\mathcal{K}_z, \mathcal{D}] = \;?$

For pure $R_2$ dephasing, a standard dissipator may be written as

$$\mathcal{D}_\phi(\rho) = \gamma_\phi(I_z \rho I_z - \frac{1}{2}\{I_z^2, \rho\}). \tag{S81}$$

This commutes with z-rotations, so $[\mathcal{K}_z, \mathcal{D}_\phi] = 0$. Therefore, coherence order $m$ is conserved: the Liouvillian remains block-diagonal in $m$, so transverse components may decay in amplitude but do not mix with other coherence orders. Symmetry is preserved. This means that $R_2$ alone does not break the $U(1)$ Killing symmetry.

For $R_1$ relaxation with energy exchange, lowering and raising amplitude-damping Lindblad operators may be written as

$$L_- = \sqrt{\gamma_\downarrow}\, I_-, \tag{S82}$$

$$L_+ = \sqrt{\gamma_\uparrow}\, I_+. \tag{S83}$$

Notably, these do not commute with $I_z$. Under rotation, we find that

$$e^{-i\theta I_z} I_- e^{i\theta I_z} = e^{-i\theta} I_- \tag{S84}$$

and likewise for $I_+$. Thus, the ladder operators $I_\pm$ are eigenoperators of the rotation generator $I_z$, with coherence order $m = \pm 1$. If rates are unequal, i.e. for finite temperature, the set of Lindblad operators is not invariant under conjugation. Then $[\mathcal{K}_z, \mathcal{D}] \neq 0$ and the Killing symmetry is broken.

Generally, because of the dissipator term, and the inseparability of $R_1$ and $R_2$ processes, the Killing vector generators do not commute with the full Liouvillian. Consequently, an open system disrupts the symmetry generated by $I_z$, and during relaxation, even two spins act as an open system and break Killing symmetry. This is where the Redfield-motivated insight that $R_2$ feeds $R_1$ and creates subdivision of system becomes relevant (vide infra).

A deeper structural interpretation utilizing the geometric framework developed here is as follows. Killing vectors correspond to conserved spin currents, and breaking them corresponds to entropy production. Therefore, spin relaxation is geometric symmetry breaking: when bath coupling is anisotropic, the dissipator does not respect spin rotation symmetry. Killing vectors in this context are the generators of spin-space isometries that survive truncation, driving, and environmental coupling.

Table S1: Summary of Hamiltonians, symmetry groups, and corresponding Killing generators relevant to NMR.

| **Hamiltonian** | **Symmetry Group** | **Killing Generator** |
|---|---|---|
| Zeeman | $U(1)$ | $I_z$ |
| Isotropic J-coupling | $SO(3)$ | $\mathbf{I}_{\text{tot}}$ |
| Secular dipolar | $U(1)$ | $I_{1z} + I_{2z}$ |
| RF resonant | $U(1)$ (axis-dependent) | frame-dependent |
| Gradient | spin $U(1)$, spatial broken | $I_z$ |
| With relaxation | generally none | symmetry broken |

## 3.2 Superoperator Killing vectors

With this intuition, we now turn to Liouville space, where Killing vectors are superoperators. This is the relevant place for curvature in spin dynamics, where spatial anisotropy generates relaxation "sinks" because both Hamiltonians and dissipators are spatially dependent in general. In this case, rotations in pulse trains do not generally commute. In Liouville space, the object with symmetry is the Liouvillian superoperator, including Hamiltonian flow and dissipation, and curvature appears as non-commutativity of generators and time ordering.

### 3.2.1 Liouville space definitions

We begin by vectorizing operators $A \mapsto |A\gg$, such that

$$|A\gg \equiv vec(A) \tag{S85}$$

and the evolution is

$$\frac{d}{dt}|\rho\gg \;=\; \mathcal{L}(t)|\rho\gg . \tag{S86}$$

Separating $\mathcal{L}(t)$ into terms for evolution under the Hamiltonian and dissipation, respectively, gives the full Liouvillian as

$$\mathcal{L}(t) = \mathcal{L}_H(t) + \mathcal{D}(t), \tag{S87}$$

with the Hamiltonian part

$$\mathcal{L}_H(t)\rho = -i\,[H(t), \rho]\,. \tag{S88}$$

In vectorized form, choosing the column-stacking convention, $\mathcal{L}_H(t)$ can be written as

$$\mathcal{L}_H(t) = -i(H(t) \otimes \mathbb{1} - \mathbb{1} \otimes H(t)^{\mathsf{T}}). \tag{S89}$$

Equivalently, it is convenient to define the left and right multiplication superoperators as

$$\mathcal{L}_A^{(L)}\rho = A\rho \tag{S90}$$

$$\mathcal{L}_A^{(R)}\rho = \rho A, \tag{S91}$$

respectively, such that $\mathcal{L}_H = -i(\mathcal{L}_H^{(L)} - \mathcal{L}_H^{(R)})$. Note that when two Killing vectors do not commute, their commutator is also a Killing vector.

### 3.2.2 Killing vectors in Liouville space are symmetry generators of $\mathcal{L}$

A continuous symmetry generated by a superoperator $\mathcal{K}$ is

$$[\mathcal{K},\, \mathcal{L}] = 0 \tag{S92}$$

for time-independent $\mathcal{L}$. More generally, for time-dependent driving, there must either be (i) instantaneous symmetry,

$$[\mathcal{K}, \mathcal{L}(t)] = 0 \tag{S93}$$

for all $t$, or (ii) Floquet symmetry,

$$[\mathcal{K}, \mathcal{U}_F] = 0 \tag{S94}$$

for the one-period propagator $\mathcal{U}_F = \mathcal{T} e^{\int_0^T \mathcal{L}(t)\,dt}$. The natural spin rotation generators in Liouville space come from adjoint action of an operator $Q$,

$$\mathcal{K}_Q\rho \equiv -i[Q, \rho], \tag{S95}$$

where

$$\mathcal{K}_Q = -i\left(Q \otimes \mathbb{1} - \mathbb{1} \otimes Q^{\mathsf{T}}\right). \tag{S96}$$

This is the Liouville space version of the familiar $U(1)/SU(2)$ generators.

### 3.2.3 Condition for a generator to be a symmetry when dissipation is present

For a generator to correspond to a symmetry with dissipation, it must commute with the Liouvillian, $[\mathcal{K}_Q, \mathcal{L}] = 0$. This can be broken into two commutators,

$$\left[\mathcal{K}_Q, \mathcal{L}_H\right] + \left[\mathcal{K}_Q, \mathcal{D}\right] = 0. \tag{S97}$$

The Hamiltonian part $\mathcal{L}_H$ typically commutes when $[Q, H] = 0$ because commutators of commutator superoperators act like commutators of operators: $[ad_Q, ad_H] = ad_{[Q,H]}$. For the dissipator part, using a general GKSL dissipator,

$$\mathcal{D}\rho = \sum_\mu \gamma_\mu \left(L_\mu \rho L_\mu^\dagger - \frac{1}{2}\left\{L_\mu^\dagger L_\mu, \rho\right\}\right), \tag{S98}$$

a sufficient covariance condition under the unitary $U(\theta) = e^{-i\theta Q}$ is:

$$U(\theta)\mathcal{D}(\rho)U(\theta)^\dagger = \mathcal{D}(U(\theta)\rho U\,(\theta)^\dagger) \qquad \forall\theta. \tag{S99}$$

This is equivalent, infinitesimally, to $[\mathcal{K}_Q, \mathcal{D}] = 0$.

In terms of Lindblad operators, this typically requires that the set $\{L_\mu\}$ transforms as a representation under $U(\theta)$ (closed under conjugation), e.g.

$$U(\theta) L_\mu U(\theta)^\dagger = \sum_\nu R_{\mu\nu}(\theta) L_\nu, \tag{S100}$$

with rates compatible with this mixing (degeneracies and equal $\gamma_\mu$ weighting inside each multiplet). This is the context in which anisotropy breaks symmetry: orientation-dependent couplings select preferred axes, so the Lindblad set is not covariant. This leads to $[\mathcal{K}_Q, \mathcal{D}] \neq 0$, which generates relaxation "sinks" in operator space.

#### 3.2.4 Anisotropic dipolar Hamiltonians and dissipators

For secular dipolar couplings in high field, the Hamiltonian depends on bond orientation:

$$H_D(\Omega) \propto \sum_{ij} b_{ij}(\Omega) \left(2 I_{iz} I_{jz} - I_{ix} I_{jx} - I_{iy} I_{jy}\right), \tag{S101}$$

with $b_{ij}(\Omega) \propto (1 - 3\cos^2\theta_{ij})/r_{ij}^3$. The corresponding Liouvillian superoperator is an ensemble of such generators,

$$\mathcal{L}(t, \Omega) = -i\, ad_{H(t,\Omega)} + \mathcal{D}(t, \Omega). \tag{S102}$$

Typically, NMR involves orientation averaging or stochastic modulation, whereby

$$\bar{\mathcal{U}}(t) = \langle \mathcal{T} \exp \left( \int_0^t \mathcal{L}(t', \Omega(t')) dt' \right) \rangle, \tag{S103}$$

and the effective reduced generator $\mathcal{L}_{\text{eff}}$ acquires dissipative terms from the fluctuations of the anisotropic pieces (e.g., in terms of the Redfield cumulant expansion). In this language, spatial anisotropy generates sinks when anisotropic components fail to commute across time and/or orientation and, after averaging, appear as non-Hamiltonian decay pathways.

#### 3.2.5 Pulse trains as non-commuting rotations that read out curvature in Liouville space

For piecewise-constant segments of free evolution interspersed with pulses, there is a sequence of Liouvillians $\mathcal{L}_1, \mathcal{L}_2, \ldots$ and the net propagator is a time-ordered product,

$$\mathcal{U}(T) = e^{\Delta t_n \mathcal{L}_n} \cdots e^{\Delta t_2 \mathcal{L}_2} e^{\Delta t_1 \mathcal{L}_1}. \tag{S104}$$

Non-commutation of the sequential Liouvillians $[\mathcal{L}_i, \mathcal{L}_j] \neq 0$ forces this time ordering. The effective generator from a Magnus expansion is

$$\mathcal{L}_{\text{eff}} = \frac{1}{T} \left( \int_0^T \mathcal{L}(t)\, dt + \frac{1}{2} \int_0^T dt_1 \int_0^{t_1} [\mathcal{L}(t_1), \mathcal{L}(t_2)]\, dt_2 + \cdots \right). \tag{S105}$$

Here, the second term is the algebraic representation of superoperator curvature built from commutators of generators. With dissipation, these commutators include cross terms such as $[\mathcal{L}_H, \mathcal{D}]$, which means, intuitively, that pulses "reshuffle" what decays.

### 3.2.6 Lie algebra closure statement

Let superoperators $\mathcal{K}$ be Killing vectors satisfying $[\mathcal{K}, \mathcal{L}] = 0$. If $[\mathcal{K}_1, \mathcal{L}] = 0$ and $[\mathcal{K}_2, \mathcal{L}] = 0$, then their commutator is also a symmetry: $[[\mathcal{K}_1, \mathcal{K}_2], \mathcal{L}] = 0$. This can be directly proven via the Jacobi identity,

$$[[\mathcal{K}_1, \mathcal{K}_2], \mathcal{L}] = [\mathcal{K}_1, [\mathcal{K}_2, \mathcal{L}]] - [\mathcal{K}_2, [\mathcal{K}_1, \mathcal{L}]] = 0 - 0 = 0. \tag{S106}$$

Thus, the symmetry generators form a Lie algebra inside $End(\mathcal{H} \otimes \mathcal{H}^*)$. One relevant subtlety for pulse trains is that if $\mathcal{L}(t)$ changes in time, it is required that $[\mathcal{K}, \mathcal{L}(t)] = 0$ for all $t$ (instantaneously) or $[\mathcal{K}, \mathcal{U}_F] = 0$ (Floquet). In many pulse sequences, $\mathcal{K}$ does not commute with the $\mathcal{L}_i$ of each segment, but a toggling frame-transformed symmetry can commute with the effective $\mathcal{L}_{\text{eff}}$ (average Hamiltonian/average Liouvillian symmetry).

### 3.2.7 Example in Liouville space: $U(1)$ about $z$ and coherence orders

Take $Q = I_z$. Then

$$\mathcal{K}_z \rho = -i\,[I_z, \rho]\,. \tag{S107}$$

Operators with definite coherence order $m$ satisfy $[I_z, A] = mA$, so in Liouville space:

$$\mathcal{K}_z \mid A \gg = -im \mid A \gg\,. \tag{S108}$$

Thus, $\mathcal{K}_z$ block-diagonalizes Liouville space by coherence order. Zeeman and secular dipolar coupling terms preserve this $U(1)$ symmetry, so $[\mathcal{K}_z, \mathcal{L}_H] = 0$. Many dissipators respect this symmetry too, particularly pure dephasing in $z$ that keeps coherence order blocks, although it damps $m \neq 0$. Anisotropic relaxation mechanisms that mix coherence orders correspond to $[\mathcal{K}_z, \mathcal{D}] \neq 0$.

## 3.3 Pulse train toggling as a connection in Liouville space

The next step to develop the picture of curvature in spin dynamics is to express pulse train toggling as a Liouville space connection $\mathcal{A}(t)$ generated by the control unitary superoperators, and to identify the Magnus commutator term $[\mathcal{L}(t_1), \mathcal{L}(t_2)]$ as the curvature governing the resulting holonomy, which plays the role of the associated two-form. This provides a framework for tracking how anisotropy-dependent dissipators $\mathcal{D}(t, \Omega)$ generate dissipative curvature, manifesting as non-unitary holonomy. To formalize this framework, below we start by (1) identifying a connection term by expressing the toggling frame in Liouville space via a gauge transformation, (2) relating curvature to noncommutativity and time-ordering of successive pulses in Magnus expansion form, and (3) formalizing the idea of curvature sinks via anisotropic, dissipative stochastic modulation.

### 3.3.1 Toggling frame in Liouville space via a gauge transformation

Begin with the evolution of spin density as

$$\dot{\rho} = \mathcal{L}(t)\rho, \tag{S109}$$

where

$$\mathcal{L}(t) = -i\, ad_{H(t)} + \mathcal{D}(t), \tag{S110}$$

and $ad_H(\rho) = [\mathcal{H}, \rho]$ represents the adjoint action of $\mathcal{H}$ on $\rho$. (In Hilbert space, $\mathcal{H}$ generates rotations; in Liouville space, $ad_H$ generates rotations of operators via commutation.) Let $U_c(t)$ be the control unitary superoperator in the pulse train frame, denoted with a tilde, and define the toggling-frame density operator as

$$\tilde{\rho}(t) = U_c^\dagger(t)\,\rho(t)\,U_c(t). \tag{S111}$$

Differentiating yields

$$\dot{\tilde{\rho}} = U_c^\dagger \dot{\rho} U_c + \dot{U}_c^\dagger \rho U_c + U_c^\dagger \rho \dot{U}_c. \tag{S112}$$

The control Hamiltonian can now be written via

$$\dot{U}_c(t) = -iH_c(t)U_c(t) \tag{S113}$$

$$\dot{U}_c^\dagger = +iU_c^\dagger H_c(t). \tag{S114}$$

By substitution, the final two terms in the toggling frame density evolution become $-i[\,\widetilde{H}_c, \tilde{\rho}\,]$ with $\widetilde{H}_c = U_c^\dagger H_c U_c$, i.e. a commutator superoperator. To express this in Liouville space via a gauge transformation, vectorize the spin density operators as $|\rho \gg$. The transformation $\tilde{\rho} = U_c^\dagger \rho U_c$then becomes

$$|\tilde{\rho} \gg = \mathcal{G}(t)|\rho \gg, \tag{S115}$$

where

$$\mathcal{G}(t) \equiv U_c^*(t) \otimes U_c^\dagger(t). \tag{S116}$$

Equivalently, $|\rho \gg = \mathcal{G}^{-1}|\tilde{\rho} \gg$ with $\mathcal{G}^{-1} = U_c^T \otimes U_c$. Differentiating now gives

$$\frac{d}{dt}\left|\tilde{\rho} \gg\right. = \dot{\mathcal{G}}\left|\,\rho \gg\right. + \mathcal{G}\left|\dot{\rho} \gg\right. = \left(\dot{\mathcal{G}}\mathcal{G}^{-1} + \mathcal{G}\mathcal{L}\mathcal{G}^{-1}\right)\left|\,\tilde{\rho} \gg\right. . \tag{S117}$$

The transformed Liouvillian is therefore

$$\widetilde{\mathcal{L}}(t) = \mathcal{G}(t)\mathcal{L}(t)\mathcal{G}^{-1}(t) + \dot{\mathcal{G}}(t)\mathcal{G}^{-1}(t) \tag{S118}$$

where the second term represents the connection. Formally, the Liouville-space connection is defined as the matrix-valued one-form

$$\mathcal{A}(t) \equiv -\dot{\mathcal{G}}(t)\mathcal{G}^{-1}(t) \tag{S119}$$

such that

$$\widetilde{\mathcal{L}}(t) = \mathcal{G}\mathcal{L}\mathcal{G}^{-1} - \mathcal{A}(t). \tag{S120}$$

Note that this is the same algebra as gauge theory, where $\mathcal{G}(t)$ is a gauge transform and $\mathcal{A}(t)$ is the induced gauge potential. We can make $\mathcal{A}$ explicit using $\mathcal{G} = U^* \otimes U^\dagger$:

$$\dot{\mathcal{G}}\mathcal{G}^{-1} = \left(\dot{U}^* U^T\right) \otimes \mathbb{1} + \mathbb{1} \otimes \left(\dot{U}^\dagger U\right). \tag{S121}$$

Because $\dot{U}U^\dagger = -iH_c$ and $\dot{U}^\dagger U = +iU^\dagger H_c U$, this becomes a commutator superoperator,

$$\dot{\mathcal{G}}\mathcal{G}^{-1} = +i\left(\widetilde{H}_c^* \otimes \mathbb{1} - \mathbb{1} \otimes \widetilde{H}_c\right) \equiv -i\,ad_{\widetilde{H}_c} \tag{S122}$$

in Liouville form. Thus, $-\mathcal{A}(t)$ is equivalent to the control commutator term that is familiar from toggling frames. The key point here is that in Liouville space, the pulse train is a gauge choice, and the control contributes a connection $\mathcal{A}(t)$.

### 3.3.2 Curvature as the obstacle to commutation under parallel transport, via the Magnus expansion

The time evolution propagator acts on an initial density of states as

$$|\rho(T) \gg = \mathcal{U}(T)|\rho(0) \gg, \tag{S123}$$

and is defined as the time-integrated exponential of the Liouvillian:

$$\mathcal{U}(T) = \mathcal{T} \exp\left(\int_0^T \mathcal{L}(t)dt\right). \tag{S124}$$

Time ordering is necessary because $[\mathcal{L}(t_1), \mathcal{L}(t_2)] \neq 0$. The Magnus expansion of the propagator around $\mathcal{L}(t)$ gives

$$\mathcal{U}(T) = \exp\left(\Omega(T)\right) \tag{S125}$$

with

$$\Omega_1 = \int_0^T \mathcal{L}(t)\, dt, \Omega_2 = \frac{1}{2}\int_0^T dt_1 \int_0^{t_1} dt_2\, \left[\mathcal{L}\left(t_1\right), \mathcal{L}\left(t_2\right)\right], \tag{S126}$$

and higher nested commutators for successive terms. Algebraically, the $\Omega_2$ term encodes the curvature; it vanishes if the generator commutes with itself across time.

In gauge theory, parallel transport of a vector $v$ is defined by $\dot{v} = -\mathcal{A}(t)v$, with the solution $v(T) = \mathcal{P}e^{-\int \mathcal{A}dt}v(0)$. The obstacle to path-independence is curvature. In spin dynamics, the same obstacle appears through second order and higher commutator terms in Magnus expansion of the propagator. Here, $\Omega_2$ is the leading order holonomy induced by the curvature, and the Liouville space connection $\mathcal{A}(t)$ is supplied by pulse control.

For an elementary example, consider a two-step loop composed of two short time steps $\delta t$ at $t_1$ and $t_2$. Then the propagator for forward order becomes

$$e^{\delta t\, \mathcal{L}(t_1)} e^{\delta t\, \mathcal{L}(t_2)} = e^{\delta t(\mathcal{L}_1 + \mathcal{L}_2) + \frac{\delta t^2}{2}[\mathcal{L}_1, \mathcal{L}_2] + O(\delta t^3)} \tag{S127}$$

while the reverse order has

$$e^{\delta t\, \mathcal{L}(t_2)} e^{\delta t\, \mathcal{L}(t_1)} = e^{\delta t(\mathcal{L}_1 + \mathcal{L}_2) + \frac{\delta t^2}{2}[\mathcal{L}_2, \mathcal{L}_1] + O(\delta t^3)} = e^{\delta t(\mathcal{L}_1 + \mathcal{L}_2) - \frac{\delta t^2}{2}[\mathcal{L}_1, \mathcal{L}_2] + O(\delta t^3)} \tag{S128}$$

using the Baker–Campbell–Hausdorff formula to demonstrate that curvature manifests as holonomy. This shows that for even the simplest possible loop, the error is governed by the commutator.

### 3.3.3 Dissipative curvature: relaxation sinks and mixing from $[\mathcal{L}_H, \mathcal{D}]$ and $[\mathcal{D}(t_1), \mathcal{D}(t_2)]$

Recall that the Liouvillian contains a second term when the dissipative effects of relaxation are considered,

$$\mathcal{L}(t) = \mathcal{L}_H(t) + \mathcal{D}(t). \tag{S129}$$

Now, the commutator in $\Omega_2$ contains three kinds of terms:

$$\left[\mathcal{L}\left(t_1\right), \mathcal{L}\left(t_2\right)\right] = \left[\mathcal{L}_H\left(t_1\right), \mathcal{L}_H\left(t_2\right)\right] + \left[\mathcal{L}_H\left(t_1\right), \mathcal{D}\left(t_2\right)\right] + \left[\mathcal{D}\left(t_1\right), \mathcal{L}_H\left(t_2\right)\right] + \left[\mathcal{D}\left(t_1\right), \mathcal{D}\left(t_2\right)\right]. \tag{S130}$$

Here, the first term represents "unitary curvature" and is well described using average Hamiltonian theory and noncommuting toggling-frame Hamiltonians. The cross terms represent "unitary–dissipative curvature", where pulse control and anisotropy together reshuffle what decays and what is conserved. The last term is entirely "dissipative curvature", where two noncommuting dissipators combine to create new effective dissipative structure. Evidently, anisotropy makes $\mathcal{D}$ both orientation- and time-dependent; as a result, its non-commutation with itself and with $\mathcal{L}_H$ produces effective irreversible generators in $\mathcal{L}_{eff}$. This is the algebraic form of the statement that spatial anisotropy generates relaxation sinks.

### 3.4 Making anisotropy explicit: Ω-dependent Liouvillian and stochastic modulation

Now let the dipolar Hamiltonian depend on orientation $\Omega(t)$ to describe molecular tumbling and diffusion on a sphere:

$$\mathcal{L}(t) = -i\, ad_{H_0+H_D(\Omega(t))} + \mathcal{D}\left(\Omega(t)\right). \tag{S131}$$

Even if the microscopic dynamics are described by a Hamiltonian with only weak bath coupling, after eliminating $\Omega(t)$ for stochastic averaging, an effective generator is obtained. A standard route for doing this is to expand the ensemble-averaged propagator in a cumulant or Magnus series,

$$\bar{\mathcal{U}}(T) = \left\langle \mathcal{T} e^{\int_0^T \mathcal{L}(t)\,dt} \right\rangle \approx \exp\left(\bar{\Omega}_1 + \bar{\Omega}_2 + \cdots\right), \tag{S132}$$

where $\bar{\Omega}_2$ contains correlation functions of the fluctuating anisotropic components. Writing $\mathcal{L}(t) = \bar{\mathcal{L}} + \delta\mathcal{L}(t)$ with $\langle \delta\mathcal{L} \rangle = 0$ yields, schematically,

$$\bar{\Omega}_2 \sim \frac{1}{2}\int dt_1 \int^{t_1} dt_2 \left\langle \left[\delta\mathcal{L}\left(t_1\right), \delta\mathcal{L}\left(t_2\right)\right] \right\rangle + (\text{symmetric products in other cumulant forms}). \tag{S133}$$

The central, enlightening point here is that anisotropy combined with finite correlation times causes these integrals to be nonzero, producing an effective decay/mixing superoperator even when the instantaneous pieces appear structured (e.g., secular dipolar). This is one mechanism by which Liouville curvature generates relaxation sinks.

#### 3.4.1 Non-unitary holonomy: rotation and decay of eigenmodes around a loop

Because $\mathcal{L}$ is generally non-normal, with $\mathcal{D} \neq 0$, its right and left eigenvectors differ:

$$\mathcal{L}(t)\,|r_n(t) \gg = \lambda_n(t) |\, r_n(t) \gg, \tag{S134}$$

$$\ll l_n(t)\,| \mathcal{L}(t) = \lambda_n(t) \ll l_n(t)|\,. \tag{S135}$$

If parameters such as pulse phase, orientation, or spinor state trace a loop in the effective control/spatial/state parameter space, the adiabatic transport of modes accumulates a holonomy governed by the non-Hermitian connection

$$\mathcal{A}_{mn}(t) = \ll l_m(t) | \dot{r}_n(t) \gg\,. \tag{S136}$$

This is the Liouville space analog of the Berry connection, but it is now complex-valued. Its imaginary part gives geometric phase-like rotation of modes, and its real part gives geometric attenuation/amplification. This describes dissipative curvature in the sense of geometric holonomy: loops also generate irreversible amplitude change, not just phase.

### 3.4.2 Lie algebra closure statement in Liouville space with time-dependence

As above, define a Killing superoperator as a generator $\mathcal{K}$ satisfying

$$[\mathcal{K}, \mathcal{L}(t)] = 0 \qquad \forall t, \tag{S137}$$

or, for the Floquet version, $[\mathcal{K}, \mathcal{U}_F] = 0$. Then, if $[\mathcal{K}_1, \mathcal{L}] = 0$ and $[\mathcal{K}_2, \mathcal{L}] = 0$, the double commutator is also a symmetry:

$$[[\mathcal{K}_1, \mathcal{K}_2], \mathcal{L}] = 0 \tag{S138}$$

by the Jacobi identity. Thus, the symmetry generators of the Liouvillian form a Lie algebra. When anisotropy and/or time-dependence breaks the symmetry, what replaces it is curvature: commutators of the instantaneous generators that appear in $\Omega_2, \Omega_3, ...$, i.e. in the holonomy.

## 3.5 Symmetry breaking and curvature in terms of the covariant relaxation tensor R

In the preceding sections, we treated (i) symmetry in Liouville space as the commutation condition $[\mathcal{K}, \mathcal{L}] = 0$, and (ii) curvature as the appearance of commutators of generators in the Magnus expansion when the Liouvillian $\mathcal{L}(t)$ varies in time (e.g., across pulse segments) and therefore does not commute with itself. Further, when dissipation is present, the Liouvillian gains a term, $\mathcal{L}(t) = \mathcal{L}_H(t) + \mathcal{D}(t)$, and the symmetry condition expands to $[\mathcal{K}, \mathcal{L}_H(t)] + [\mathcal{K}, \mathcal{D}(t)] = 0$. The first term is unitary as $\mathcal{L}_H(t)$ generates reversible parallel transport; its noncommutativity across pulse segments was interpreted as geometric curvature in Liouville space. Time-independent spatial anisotropy in the relevant Hamiltonian also serves to break its symmetry.

Thus far, the dissipative term $\mathcal{D}(t)$ has been discussed primarily in terms of covariance of the Lindblad operator set. However, the same symmetry question can be expressed more universally in terms of a covariant relaxation tensor $R$, introduced in a previously developed geometric framework to unify Bloch (Hilbert state space) and Lindblad (Liouville operator space) representations of dissipation. Moreover, the covariant relaxation tensor encapsulates the result from Redfield theory that $R_1$ is inseparable from $R_2$ by combining them in a frame-independent tensorial object, akin in a sense to chemical shift anisotropy and diffusion tensors [ref relaxation tensor paper]. Physically, the off diagonal elements of $R$ are basis-dependent, but the scalar contraction is invariant. Mixing becomes physically meaningful once the basis is fixed by symmetry (coherence-order sectors etc.).

This geometric framework for relaxation replaces operator-level statements with a tensorial one that can be projected into any relevant frame or space. When $R$ is projected from a symmetric (0,2) tensor (positive semidefinite) to a (1,1) bilinear form $R^{\mu}{}_{\nu}$, it acts as relaxation operator on measurable spin coordinates $\mathbf{S}$. In component form, $\mathbf{S}$ is defined as the trace of the density operator with respect to a chosen operator basis $\{A_\mu\}$,

$$S_\mu = \mathrm{Tr}\left(\rho A_\mu\right). \tag{S139}$$

In reduced form, the dynamics may be written as

$$\dot{S}_\mu = \Omega_{\mu\nu} S_\nu - R^{\mu}{}_{\nu}\, \delta S^{\nu}, \tag{S140}$$

where $\mu, \nu$ label components of $\mathbf{S}$, $\Omega_{\mu\nu}$ represents coherent evolution (holonomy), $R^{\mu}{}_{\nu}$ is the relaxation rate operator, and the deviation from equilibrium is $\delta S^{\nu} = S^{\nu} - S^{\nu}_{eq}$. Here $R^{\mu}{}_{\nu}$ collects both decay rates and dissipative couplings between components of $\mathbf{S}$. Thus, it is the reduced coordinate representation of the irreversible part of the Liouvillian.

### 3.5.1 Symmetry condition in terms of $R$

In a coordinate representation adapted to the Liouville space symmetry generator $\mathcal{K}$, its action induces a linear operator $K^{\mu}{}_{\nu}$ on the spin coordinate vector. As before, symmetry of the full dynamics requires $[\mathcal{K}, \mathcal{L}] = 0$ and separates into two independent conditions:

$$[K, \Omega] = 0 \tag{S141}$$

$$[K, R] = 0. \tag{S142}$$

The first is geometric (holonomic invariance) and has already been discussed. The second is dissipative (contractive invariance) and is developed below.

In reduced coordinates, this implies that both the coherent and relaxation operators commute with the symmetry generator. For the relaxation part, the condition becomes

$$[K, R]^{\mu}{}_{\nu} = 0, \tag{S143}$$

where the commutator is written explicitly in components as

$$[K, R]^{\mu}{}_{\nu} = K^{\mu}{}_{\alpha} R^{\alpha}{}_{\nu} - R^{\mu}{}_{\alpha} K^{\alpha}{}_{\nu}. \tag{S144}$$

If this expression vanishes, relaxation preserves the invariant subspaces defined by $\mathcal{K}$. For example, if $\mathcal{K}$ generates axial rotations, these subspaces correspond to coherence order blocks. In that case, $R^{\mu}{}_{\nu}$ does not couple different coherence orders. However, if instead

$$[K, R]^{\mu}{}_{\nu} \neq 0, \tag{S145}$$

then relaxation couples components that transform differently under $\mathcal{K}$. In a symmetry-adapted basis, this appears as off-block-diagonal structure of $R^{\mu}{}_{\nu}$. Physically, relaxation transfers amplitude between otherwise invariant components (e.g., between different coherence orders), and the symmetry is broken by dissipation.

### 3.5.2 Relationship to entropy production

The relaxation operator also defines a nonnegative contraction of deviations from equilibrium,

$$\sigma = \delta S_{\mu} \, R^{\mu}{}_{\nu} \, \delta S^{\nu}, \tag{S146}$$

which represents the irreversible loss of distinguishability in the reduced description. When $R^{\mu}{}_{\nu}$ couples distinct $\mathcal{K}$-invariant subspaces, this contraction includes cross terms between components that would otherwise remain dynamically separated. In that sense, symmetry-breaking relaxation not only damps components but also mixes them while generating entropy.

### 3.5.3 Summary

Writing the propagator in Liouville space as

$$\mathcal{U}(T) = \mathcal{T} \exp\left( \int_0^T \mathcal{L}(t) \, dt \right) = \exp\left( \int_0^T \mathcal{L} \, dt + \frac{1}{2} \int \int_{t_2 < t_1} [\mathcal{L}(t_1), \mathcal{L}(t_2)] \, dt_2 dt_1 + \cdots \right) \tag{S147}$$

with the toggling-frame gauge transformation

$$\widetilde{\mathcal{L}}(t) = \mathcal{G} \mathcal{L}(t) \mathcal{G}^{-1} - \mathcal{A}(t), \mathcal{A}(t) \equiv -\dot{\mathcal{G}} \mathcal{G}^{-1} \tag{S148}$$

shows that dissipative curvature arises because the commutator includes the cross term

$$[\mathcal{L}_H(t_1), \mathcal{D}(t_2)], \tag{S149}$$

and the mixing term

$$[\mathcal{D}(t_1), \mathcal{D}(t_2)], \tag{S150}$$

which together generate effective sinks and relaxation exchange (in the form of $R_2$ dephasing entropy and $R_1$ energy transfer) when anisotropy makes $\mathcal{D}$ parameter-dependent and noncommuting across time.

This is more effectively and concisely communicated via the covariant relaxation tensor, $R$. The symmetry condition for dissipative dynamics can be stated directly as a commutation condition between the Liouville space symmetry generator and the relaxation operator:

$$[\mathcal{K}, R] = 0 \text{ (symmetry preserved)}, [\mathcal{K}, R] \neq 0 \text{ (symmetry broken by relaxation)}. \tag{S151}$$

This formulation makes explicit that dissipation breaks symmetry when the induced relaxation operator fails to preserve the invariant subspaces defined by $\mathcal{K}$. In this way, anisotropy and time dependence generate curvature through noncommutation of generators, while the relaxation operator determines whether that curvature acquires an irreversible, entropy-producing component.

# 4 Experimental and phenomenological setup

## 4.1 Pulse train example: deterministic mixing of $T_1$ and $T_2$

Consider an exemplary train of hard pulses separated by a repetition time $T_R$. Let the $n$th pulse have tip angle $\alpha_n$ about $\hat{x}$. Between pulses, the longitudinal magnetization evolves as

$$M_z^-(n+1) = M_0 - (M_0 - M_z^+(n))e^{-TR/T_1}. \tag{S152}$$

While immediately after the pulse

$$M_\perp^+(n) = M_z^-(n) \sin\alpha_n \cdot e^{i\varphi_n}, \quad M_z^+(n) = M_z^-(n) \cos\alpha_n. \tag{S153}$$

In the small-angle regime $\alpha_n \ll 1$, expanding $\sin\alpha \approx \alpha$ and $\cos\alpha \approx 1 - \alpha^2/2$ gives

$$M_z^+(n) \approx M_z^-(n)\left(1 - \frac{\alpha_n^2}{2}\right), \tag{S154}$$

so the pulse train itself introduces an effective longitudinal decay pathway that is purely geometric, even before environmental relaxation acts. At the same time, the observable signal

$$S_n \propto M_\perp(n) \approx \alpha_n \, M_z^-(n) \, e^{-TE/T_2} \tag{S155}$$

retains explicit transverse relaxation. However, we will see (vide infra, § 5.3) that this naïve picture is generally invalid.

While the experiment can deterministically couple longitudinal and transverse pathways through control rotations, the bath couples them through $R_{\mu\nu}$. The resulting hybrid behavior reflects the interplay of unitary mixing with dissipative projection – a situation readily captured by the tensor formulation but difficult to describe using separate $T_1$ and $T_2$ constants.

## 4.2 Multicomponent relaxation

At the operator level, evolution of multicomponent spin density can be described as a sum over Liouvillians

$$\rho(t) = \sum_m \omega_m e^{\mathcal{L}_m t} \rho(0) \tag{S156}$$

with $\omega_m$ giving the weight of each $m$th component. At the Bloch level of observables, this corresponds to a time domain signal

$$S(t) = \sum_m \omega_m \phi_m(t) S(0), \tag{S157}$$

where $\phi_m(t)$ describes the weighted exponential decay of each $m$th component, overall normalized to the magnitude of the initial signal intensity. Across these representation levels, each component of relaxation has its own generator,

$$\mathcal{L}_m = \mathcal{L}_{H,m} + \mathcal{L}_{\mathcal{D},\ m}, \tag{S158}$$

thus the observed signal arises from a weighted sum of propagators, each generated by its own $R^{(m)}$. Note that this is because each $R_1$ or $R_2$ component corresponds to a different sub-ensemble of the spin system; the signal is therefore a superposition of trajectories (*not* one trajectory with a summed rate). Therefore, for sequential time-ordered cyclic steps $k$, the decay of each component measured by a pulse sequence is

$$\phi_{seq}^{(m)} = \tau \prod_k \exp\left[\left(\mathcal{L}_{H,k} + \mathcal{L}_{\mathcal{D},\ k}\right) \Delta t_k\right], \tag{S159}$$

with $R_1$ or $R_2$ projections of the same fitted generator(s).

Overall, the measured relaxation tensor is an effective quantity $R_{eff}$ related to the underlying, covariant $R_{\mu\nu}$ by a Floquet map $M$:

$$R_{eff} = [M] R_{\mu\nu}, \tag{S160}$$

where $M$ contains the $\mathcal{L}_H$ and $\mathcal{L}_{\mathcal{D}}$ relevant for the system and the control sequence. The covariant $R_{\mu\nu}$ may thus be obtained by inverting the Floquet map on the measured $R_{eff}$,

$$R_{\mu\nu} = [M]^{-1} R_{eff}, \tag{S161}$$

or for multiple components,

$$R_{\mu\nu}^{(m)} = \sum_m [M]^{-1} R_{eff}^{(m)}. \tag{S162}$$

## 4.3 The necessity of a full tensor: when ordering matters

The diagnostic condition is

$$e^{\mathcal{L}_2 \Delta t}\, e^{\mathcal{L}_1 \Delta t} \neq e^{\mathcal{L}_1 \Delta t}\, e^{\mathcal{L}_2 \Delta t} \tag{S163}$$

where $\mathcal{L}_k = \mathcal{L}_H(t_k) + \mathcal{L}_D(t_k)$, and the lefthand side represents one phase ordering, compared to the same content with different ordering on the righthand side. If an experiment explores multiple axes in a way that makes $[\mathcal{L}_H, \mathcal{L}_D] \neq 0$ (i.e., through detuning that results in a tilted effective field, anisotropic couplings, quasi-continuous driving, etc.), then a scalar kernel model like $e^{-t/T_1} e^{-t/T_2}$ and stretched exponential modifications cease to be representation-invariant. These circumstances motivate the use of a full tensor.

This is also why pulse trains can perform effectively for AC sensing without phase cycling. These experiments often operate in regimes where the measurement objective is robust to these non-commuting aspects when the goal is a stable orbit converted to a usable sensor response, rather than inversion of the dynamics to identify a constitutive tensor. Thus far, the applications of quantum control, AC sensing, and magnetometry have largely separated their effectiveness from this inversion problem by emphasizing long-lived driven trajectories and emergent decay behavior, not first principles GKLS identification. The full tensor inversion offers a way to identify the constitutive relaxation tensor for an open, driven spin medium, not just an envelope fit.

## 4.4 The case of inseparable $R_1$–$R_2$ mixing

As developed in the main text, in a detuned, anisotropic system that is driven continually (or quasi-continually), spins experience an instantaneous effective field $B_{eff}(t)$. A dressed basis follows this tilted axis, but the laboratory frame readout basis does not. When the basis rotation is time-dependent, transforming a diagonal dressed frame relaxation operator back to the lab frame produces off-diagonal lab frame components,

$$R_{\text{lab}}(t) = \Lambda(t)\, R_{\text{dressed}}\, \Lambda(t)^{-1}. \tag{S164}$$

Here, even if $R_{\text{dressed}}$ is just $T_{1\rho}, T_{2\rho}$ at any instant, the lab frame object can appear to have mixing because the basis is moving. This is a deterministic route to inseparability, where the simplistic approximation in §4.1 becomes immediately invalid, and it instead aligns with the view of spin evolution as parallel transport under a connection that can sometimes be hidden by the rotating frame.

Caution is warranted when fitting an arbitrary constant 3×3 $R$ while the true lab frame $R(t)$ is being rotated by $\Lambda(t)$. The fit can conceal basis motion physics as static off-diagonals; this can still provide a useful effective tensor, but it should be labeled as such, e.g., a time-averaged effective relaxation tensor over one Floquet period.

## 4.5 $S_x$, $S_y$ trajectories as a diagnostic for generator commutation

The shape of $S_x, S_y$ trajectories during a pulse train comes directly from the algebra of the pulse sequence. If the generators commute and the effective Hamiltonian during the pulse train is dominated by one fixed generator, the transverse magnetization evolves as

$$\dot{\mathbf{S}} = \Omega \times \mathbf{S} - R\mathbf{S}, \tag{S165}$$

which produces circular or elliptical trajectories in $S_x, S_y$. This is demonstrated in Fig. 3, where smooth nested curves result from a train of x pulses following an initial y pulse for varied transmitter frequency offset and pulse tip angle. Here each point corresponds to the magnetization after a pulse, and what is plotted is $\big(S_x(n), S_y(n)\big)$ for successive pulses $n$. In some cases, the points form clusters near the origin. This occurs where the trajectory momentarily slows in phase space and may happen when the effective rotation axis aligns partly with the magnetization vector. This effect is more noticeable with small tip angles (yellow).

When pulses do not commute, successive pulses apply rotations about different axes. Schematically, the propagator is proportional to the time-ordered product of the matrix exponentials:

$$U \propto e^{-i\theta_1 S_x}\, e^{-i\theta_2 S_y}\, e^{-i\theta_3 S_x}\, e^{-i\theta_4 S_y} \ldots \tag{S166}$$

Because spin operators do not commute with each other,

$$[S_x, S_y] = iS_z, \tag{S167}$$

each pair of pulses introduces an additional generator. Thus the Magnus expansion begins as

$$\Omega = \theta_1 S_x + \theta_2 S_y + \frac{1}{2}[\theta_1 S_x, \theta_2 S_y] + \cdots \tag{S168}$$

which produces a secondary rotation about $S_z$. This is a geometric phase term.

For a pulse sequence that alternates orthogonal axes, the train of x pulses can be replaced with cyclic winding of (x, y, -x, -y) in the positive sense, or (x, -y, -x, y) for the negative sense. Each step of the evolution alternates a rotation around $x$ with a rotation around $y$. In the transverse projection, this produces a trajectory that advances in piecewise orthogonal segments, repeated hundreds of thousands of times. When plotted continuously, this becomes a rotating square-spiral pattern as shown in Figs. 3 and 5. Here, the rotation of the square trajectory appears because the geometric phase term from the commutator produces an effective slow precession around $z$. Thus, the appearance of nested, rotating square trajectories arises from the combination of fast orthogonal kicks from $S_x, S_y$ with slow rotation from $S_z$ (the commutator). At smaller tip angles (corresponding to yellow centers in the color map), rotations are small and trajectories stay near the origin. Hence, those traces are compact and the geometric term accumulates slowly, compared to larger tip angles (shown in blue) that push the trajectories farther outward.

### 4.6 Experimental motivation for the relaxation tensor

Figures 3 and 5 in the main text visually show that the pulse train is sampling non-commuting directions in Liouville space. This means the experiment probes different tensor components of relaxation. While the commuting sequence isolates mostly diagonal components, leading to simple trajectories, the phase-wound sequences mix components and lead to more complex, geometric trajectories. Moreover, careful inspection of the geometric trajectories in Fig. 3 reveals a subtle but important feature: the corners of the square structures in phase space are slightly rounded. It is most likely that the rounding appears from relaxation during the pulse train, rotating frame offsets, and finite pulse width effects. Thus, the geometry of the spin trajectories is shaped by the intrinsic relaxation tensor as well as the pulse sequence engineering.

## 5 Geometric and algebraic interpretation of observed trajectories

### 5.1 Baker–Campbell–Hausdorff structure of SU(2)

The square-spiral trajectory patterns produced by pulse train phase winding are a direct geometric signature of non-Abelian evolution. Mathematically, the pulse sequence is a series of rotations in SU(2)

$$U = e^{-i\theta_1 S_x} e^{-i\theta_2 S_y} e^{-i\theta_3 S_x} e^{-i\theta_4 S_y} \dots \tag{S169}$$

where each pulse is a group element. The trajectory plotted in $S_x, S_y$ is the projection of spin density into observable space as

$$\rho(t) = U(t)\rho_0 U^\dagger(t). \tag{S170}$$

If the generators commuted, $[S_x, S_y] = 0$, then the order of pulses would not matter and the product would collapse to

$$U = e^{-i(\theta_x S_x + \theta_y S_y)}. \tag{S171}$$

The motion would simply be a rotation. However, for spinors,

$$\left[S_x, S_y\right] = iS_z \tag{S172}$$

and the algebra does not close in two dimensions. Therefore, every time $S_x$ and $S_y$ are combined, $S_z$ is generated. When two rotations are combined as $e^A e^B$, the Baker–Campbell–Hausdorff (BCH) expansion gives

$$e^A e^B = \exp\left(A + B + \frac{1}{2}[A, B] + \frac{1}{12}\left[A, [A, B]\right] - \frac{1}{12}\left[B, [A, B]\right] + \cdots\right). \tag{S173}$$

Using this, the effective generator becomes

$$\Omega = A + B + \frac{1}{2}[A, B] + \cdots \tag{S174}$$

Denoting pulses as $A = -i\theta S_x$ and $B = -i\theta S_y$,

$$[A, B] \propto S_z \tag{S175}$$

where the commutator term is the geometric phase generator.

## 5.2 Operator space curvature and spinor parallel transport

In gauge language, a pulse sequence is equivalent to transporting a state along a path in control space. Here the connection is

$$A = A_x S_x + A_y S_y \tag{S176}$$

and the curvature is

$$F = dA + A \wedge A. \tag{S177}$$

The wedge product in the second term produces

$$A_x A_y \left[S_x, S_y\right], \tag{S178}$$

which gives $S_z$. Physically, this term appears in the curvature when generators fail to commute.

## 5.3 Dissipative evolution as a covariant transport problem in Liouville space

The trajectories shown in Fig. 3 of the main text arise from approximating a discrete parallel transport loop in spinor space. This is an experimental visualization of BCH curvature in spin dynamics. In the phase-wound experimental case, the square spiral trajectories demonstrate the curvature of the control connection acting on the density operator manifold in the slow rotation of the square lattice in phase space. Here the phase cycle closes a loop in four steps, with the pulse phase incremented by 90° between successive pulses:

$$U_x(\theta),\ U_y(\theta),\ U_{-x}(\theta),\ U_{-y}(\theta), \tag{S179}$$

where $\theta$ is the tip angle set by the pulse width. This control sequence imposes a $C_4$ symmetry in SU(2), and the four pulse phases complete the control loop, producing the square structure. Spin trajectories are constrained with respect to this fourfold rotational symmetry.

If the generators commuted, the square would close exactly. Here, however, geometric phase accumulation and transmitter frequency offsets break perfect closure. Both these effects introduce a small extra rotation per cycle, which twists the square slightly over time and produces the spiral pattern in the overlays. The overlays plotted in Fig. 3 show the path in phase space twisting because of non-commutativity.

When the cycle is repeated thousands to millions of times, the slow geometric rotation per cycle accumulates into a large effect, which gives rise to nested squares and contracting spiral rotation. Each pulse moves along one generator of the Lie algebra, then another, and another, etc. The failure of the loop to close is the curvature term $[A, B]$ from the BCH expansion, and the rotating square trajectory is a visualization of the commutator. More generally, this is illustrated in Fig. 1 with non-commuting generators that fail to close a parallelogram in operator space.

## 5.4 Holonomy in spin dynamics

Figures 3 and 5 in the main text show the holonomy of the control loop acting on the spin state. Each cycle traces a loop in control space and because the generators do not commute, the loop produces a small residual rotation. This residual rotation is a manifestation of the geometric phase and operator curvature. The way the arcs cluster suggests the effective rotation axis per cycle is not exactly along z, but slightly tilted. This tilt can contain information about the off-diagonal components of the relaxation tensor we seek to measure.

It follows that if the pulse cycle were changed, for example, to three axes instead of two, the trajectories would become triangular spirals. If we used five, six, or more symmetric directions, we would expect to observe pentagonal, hexagonal, etc. patterns. Hence the polygon that appears directly encodes the symmetry of the control algebra. Note that many NMR experiments hide such structure because the signal is averaged or projected onto a fixed axis.

## 5.5 Gauge potential and geometric Hamiltonian

Geometric phase arises from a gauge structure that can be represented in the Hamiltonian

$$H_{geo}(t) = \hbar \frac{\partial \boldsymbol{\lambda}}{\partial t} \cdot \mathbf{A}_\lambda \tag{S180}$$

where $\mathbf{A}_\lambda$ is a gauge vector potential defined by

$$\mathbf{A}_\lambda \equiv -iU_{rf}^\dagger \frac{\partial U_{rf}}{\partial \boldsymbol{\lambda}}, \tag{S181}$$

with $\lambda$ a coordinate on the parameter manifold that defines the spinor path. In NMR, this gauge term arises from the time dependence of the rf rotation axis; its integral over one winding cycle corresponds to the experimentally observed geometric phase increment per echo. For spin-½, the matrix-valued connection of $\mathbf{A}_\lambda$ can be expanded on the $SU(2)$ generators as $-iU_{rf}^\dagger \dot{U}_{rf} = \mathbf{a}(t) \cdot \mathbf{S}$, with $\mathbf{a}(t)$ the generator coefficients, such that $H_{geo}(t) = \hbar \mathbf{a}(t) \cdot \mathbf{S}$. More generally, phase in NMR is associated with gradients in electromagnetic gauge fields imposed by $B_0$ or $B_1$ fields.[5] Projecting $\mathbf{A}_\lambda$ onto a locally followed eigenstate maps this term to an experimental observable; its time integral yields the geometric phase.

For a cyclic evolution of a quantum state, $|\psi(t)\rangle = e^{i\alpha}|\psi(0)\rangle$, the accumulated geometric phase

over time $T$ is

$$\begin{aligned}\phi_{geo} &= \alpha - \frac{1}{\hbar}\int_0^T \langle\psi(t)|H(t)|\psi(t)\rangle\, dt \\ &= \arg\langle\psi(0)|\psi(T)\rangle - Im\int_0^T \left\langle \psi(t)\middle|\frac{\partial\psi(t)}{\partial t}\right\rangle dt\end{aligned} \tag{S182}$$

In the present experiments, the cyclic evolution corresponds to one complete phase-winding cycle of the pulse train, such that the accumulated geometric phase manifests directly as a systematic phase shift between successive pulses.

With $U(1)$ gauge structure, the connection $\mathbf{A}_\lambda = i\langle n(\lambda)|\partial_\lambda n(\lambda)\rangle$ acts as vector potential in control parameter space, and the curvature $\mathbf{F}_{\lambda\lambda'} = \partial_\lambda \mathbf{A}_{\lambda'} - \partial_{\lambda'}\mathbf{A}_\lambda$ corresponds to the associated magnetic field[5, 9]. Here $\lambda$ and $\lambda'$ label coordinates on the control manifold (e.g., $\lambda = \theta$, $\lambda' = \phi$ for the Bloch sphere). More generally, for a matrix-valued gauge connection the curvature is given in component form by $F_{\lambda\lambda'} = \partial_\lambda A_{\lambda'} - \partial_{\lambda'} A_\lambda + [A_\lambda, A_{\lambda'}]$, or, in exterior calculus form, $F = \mathbf{d}A + A \wedge A$.[5]

In the present case, where the Berry connection is projected onto a single instantaneous eigenstate, the connection is Abelian and the commutator term vanishes. This reduction mirrors electromagnetism, where the U(1) gauge field is Abelian and $[A_\lambda, A_{\lambda'}] = 0$. For non-Abelian gauge fields or degenerate subspaces, however, the commutator term plays an essential role.[9] The geometric phase represents the holonomy acquired by parallel transport along a closed path, which is directly related to operator space curvature

$$\phi_{\text{geo}} = \oint \mathbf{A}_\lambda d\lambda = \iint \mathbf{F} d\mathbf{S}. \tag{S183}$$

By Stokes' theorem, the line integral of the gauge potential around the closed path is equivalent to the surface integral of its curvature over the enclosed area. Both are gauge-invariant modulo $2\pi$. Here, $\lambda$ denotes the rf phase, so $\pm$-winding reverses the loop orientation from $\phi_{\text{geo}} \to -\phi_{\text{geo}}$.

## 5.6 Geometric phase does not interact with relaxation

Upon simultaneous measurement of geometric phase (holonomy) and relaxation, one might question why the accumulated geometric phase continues to grow linearly for a fixed phase-winding loop even as relaxation causes the spin trajectory to spiral up the Bloch sphere, such that the transverse area becomes progressively smaller. The answer is that the measured geometric phase is set by the control loop, here determined by the phase winding, and not by the instantaneous size of the Bloch vector. Relaxation shrinks the transverse projection of the Bloch vector, but it does not change the mathematical SU(2) connection $A$ that is integrated to give rise to geometric phase. The salient point is that this integral is in the control space of operators, i.e. Liouville space, not in the space of magnitudes of magnetization vectors. Hence, per cycle, geometric phase accumulates in fixed increments

$$\Delta\phi_{geo} = -m_s\Omega \tag{S184}$$

for a pure spinor eigenstate $|m_s\rangle$ traversing a loop enclosing solid angle $\Omega$.[10, 11, 12, 13]

In bulk NMR, the detected phase is tied to spin coherences arising from off-diagonal density matrix elements $\rho_{m_s m_s'}$ that acquire the geometric phase difference

$$\Delta\phi_{m_s m_s'} = \phi_{m_s} - \phi_{m_s'} = -(m_s - m_s')\Omega. \tag{S185}$$

For $m_s = \pm\frac{1}{2}$, single-quantum coherences are $(m_s - m_s^{'}) = 1$ and $\Delta\phi_{m_s m_s^{'}} = -\Omega$, consistent with the experimentally extracted phase increment per winding cycle. For a loop repeated $n$ times, the total accumulated geometric phase is thus $\phi_{geo} = -n\Omega$.

Conversely, relaxation acts in the Bloch space of magnetization vectors – not in the control space of operators. The basic action of relaxation is to rescale the amplitude of magnetization states (i.e., schematically as $M(t) \approx e^{-Rt} M_{unitary}(0)$, omitting constants and more complex kernels here for simplicity). This is a dissipative contraction, not a change of the control path or projective trajectory. Notably, geometric phase depends only on the trajectory of a state, not its norm. Hence the holonomy continues to accrue identically as long as the phase winding loop is maintained, regardless of relaxation (until SNR has dropped to the point that there is no more signal to record; Fig. S1).

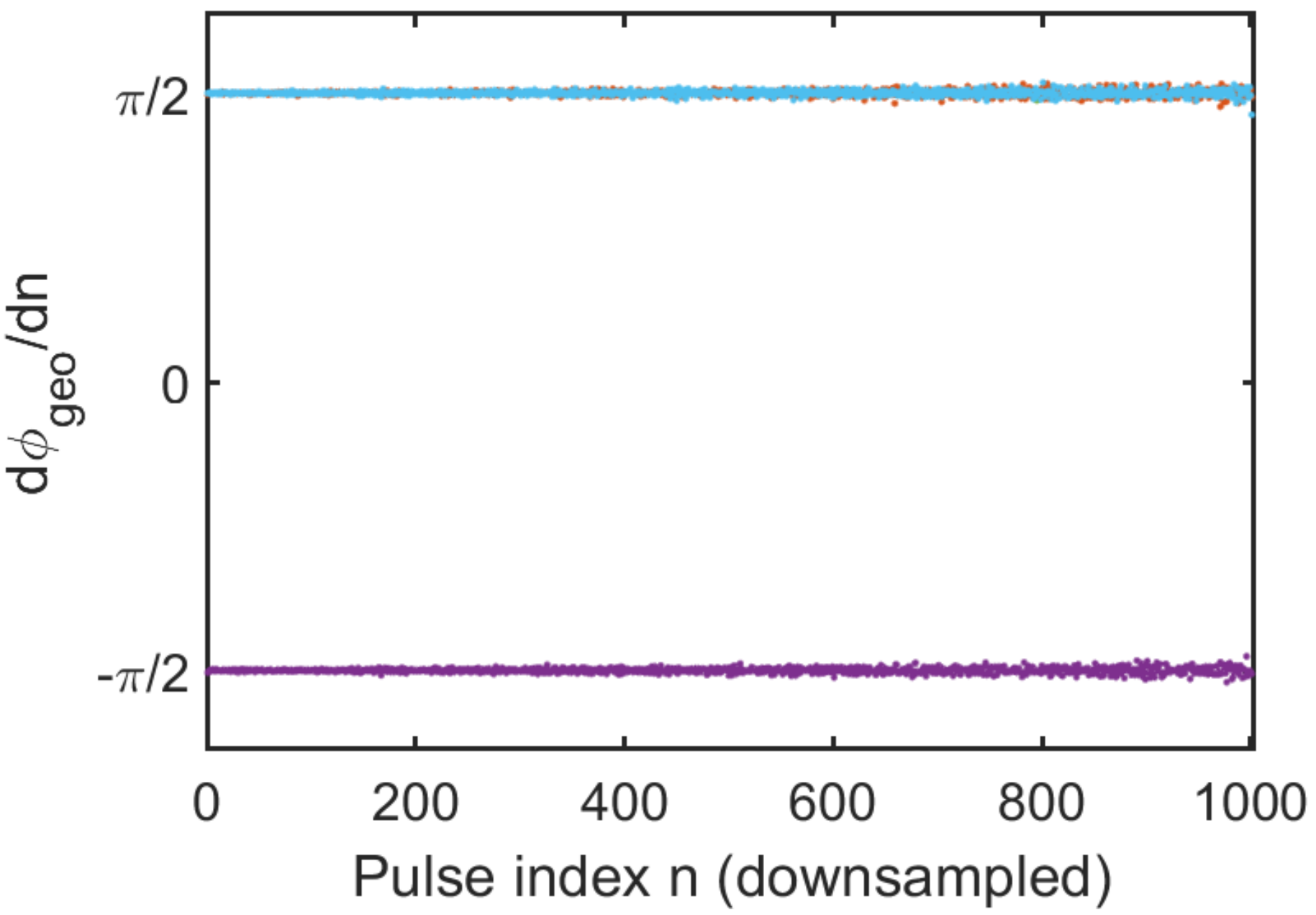

Fig. S1: **Derivative of geometric phase with respect to pulse index.** Data were downsampled from ~1 million pulses for processing, showing consistent accumulation in increments of $+\pi/2$ and $-\pi/2$ for positive and negative winding, respectively, across offset frequencies and pulse widths.

This can also be understood from the superoperator perspective, where the Liouvillian includes Hamiltonian and dissipative terms,

$$\mathcal{L}(t) = \mathcal{L}_H(t) + \mathcal{D}(t). \tag{S186}$$

Here the Hamiltonian term $\mathcal{L}_H(t)$ includes the rf pulse control and hence generates the holonomic connection that is measured as geometric phase. Relaxation is separate and contained in $\mathcal{D}(t)$.

## 5.7 Stroboscopic mapping of the pulse cycle

*Case 1: Isotropic relaxation tensor*

Suppose the effective evolution per pulse cycle were purely unitary with isotropic transverse decay. Then the effective generator would look like

$$\mathcal{L} = \Omega_x \widehat{S}_x + \Omega_y \widehat{S}_y + \Omega_z \widehat{S}_z - R_2 \left( S_x^2 + S_y^2 \right). \tag{S187}$$

In the $S_x, S_y$ projection, this would produce rotation and radial contraction. The trajectory would look like a circle spiraling toward the origin. Importantly, the spiral would stay symmetric.

*Case 2: Relaxation tensor anisotropy*

The arcs of the $S_x, S_y$ trajectories in Fig. 3 of the main text are not symmetric. In each of the four lobes, central parts of the arcs are vertical clusters while outer regions are stretched diagonally. In other words, the evolution speed is different depending on direction in the plane, which strongly hints that the effective dissipative operator is anisotropic.

Treating dissipation as a tensor,

$$R = \begin{pmatrix} R_{xx} & R_{xy} & R_{xz} \\ R_{yx} & R_{yy} & R_{yz} \\ R_{zx} & R_{zy} & R_{zz} \end{pmatrix} \tag{S188}$$

acting in Liouville space. If only diagonal terms existed, the decay rates along $x$ and $y$ directions would be the same. If off-diagonal components exist (for example $R_{xy}$) then decay couples the axes and causes shearing of the trajectories.

During each pulse segment the state evolves roughly as

$$\dot{\mathbf{S}} = \Omega \times \mathbf{S} - R\mathbf{S} \tag{S189}$$

If $R$ has cross-terms, then

$$\dot{S}_x \propto -R_{xx} S_x - R_{xy} S_y \dot{S}_y \propto -R_{yy} S_y - R_{yx} S_x \tag{S190}$$

which mixes the directions. Geometrically, instead of shrinking isotropically, the trajectory becomes tilted and stretched.

Dense vertical clusters occur when the instantaneous motion is mostly along one direction. This happens when the effective rotation axis aligns with an eigenvector of the dissipative tensor. At those points, motion in the orthogonal direction slows, so points become less dispersed. These clusters therefore hint at the principal axes of the relaxation operator.

## 5.8 Relaxation tensor tomography with geometric sequences

The commuting pulse train (here along x) samples approximately the same Liouvillian direction throughout its entirety; thus, relaxation tensor anisotropy is largely hidden. In contrast, the geometric phase-wound pulse trains reveal this effect because the sequence continually rotates the frame. As a result, successive pulses probe different projections of the tensor. In a sense, the geometric sequence is performing tomography on the tensor.

Here, this makes anisotropy appear as distorted squares, unequal arc curvature, and clustering. Together, the results visually support the behavior of relaxation as a covariant tensor that can be revealed by non-commuting pulse trains. Crucially, the pattern only appears when the control operators do not commute. The overall symmetry of the four lobes suggests the dominant anisotropy is still aligned close to the lab $x, y$ axes. If the tensor were strongly rotated, the lobes would appear skewed relative to the axes. Thus the principal frame may already be close to the detection frame, which would simplify extraction of $R$ in the following section.

# 6 Fitting procedure for $R$: parameter estimation and Floquet inversion

The purpose of this approach is to extract and compare sequence-dependent effective projections of a covariant relaxation tensor $R_{\mu\nu}$, and to establish a framework for reconstructing the underlying tensor from control-dependent measurements.

## 6.1 The objects to fit: effective generators and projected tensors

In the present experiments, the measured observables are the transverse magnetization components $S_x(n)$ and $S_y(n)$, sampled stroboscopically at each pulse index $n$. These trajectories encode the combined effects of coherent control, detuning, and dissipation under a given pulse sequence. Rather than attempting to directly fit scalar decay envelopes or independent $R_1$ and $R_2$ processes, we treat the dynamics as arising from an effective linear generator acting on the transverse subspace. Specifically, for each dataset, we model the evolution as

$$\frac{d}{dn}\begin{pmatrix} S_x \\ S_y \end{pmatrix} = A_{eff}\begin{pmatrix} S_x \\ S_y \end{pmatrix}, \tag{S191}$$

where $A_{eff}$ is a $2 \times 2$ effective generator describing the coarse-grained, cycle-averaged evolution under the pulse train. This generator captures both coherent and dissipative contributions to the transverse dynamics.

We decompose $A_{eff}$ into symmetric and antisymmetric parts,

$$A_{eff} = -R_\perp + \Omega, \tag{S192}$$

where

$$R_\perp = -\frac{1}{2}\left(A_{eff} + A_{eff}^T\right) \tag{S193}$$

is the symmetric dissipative component, and $\Omega = \frac{1}{2}(A_{eff} - A_{eff}^T)$ represents coherent rotation in the transverse plane. The matrix $R_\perp$ defines an effective transverse relaxation tensor,

$$R_\perp = \begin{pmatrix} R_{11} & R_{12} \\ R_{21} & R_{22} \end{pmatrix}, \tag{S194}$$

whose components are reported in Fig. 4 of the main text.

It is important to emphasize that $R_\perp$ is not the full covariant relaxation tensor $R_{\mu\nu}$, but rather a sequence-dependent effective projection onto the observed $(S_x, S_y)$ subspace. In the presence of detuning, continual driving, and noncommuting pulse sequences, the underlying relaxation dynamics are transformed by the control into a cycle-averaged effective generator. The fitted $R_\perp$ therefore reflects both intrinsic dissipation and the manner in which it is sampled and mixed by the pulse sequence.

In particular, for noncommuting (phase-wound) pulse trains, the effective generator corresponds to a Floquet or cycle-averaged object obtained from the ordered product of coherent and dissipative evolution over one pulse cycle. As a result, the extracted tensor components depend on the control protocol (e.g., winding, offset, and tip angle), even though they arise from a single underlying dissipative structure. Commuting pulse trains provide a near-diagonal reference projection, while noncommuting trains expose off-diagonal structure through ordering sensitivity and Liouville-space holonomy.

Thus, the quantities $R_{11}, R_{12}, R_{21}, R_{22}$ reported here should be interpreted as effective transverse projections of a covariant relaxation tensor, measured under specific control histories. Establishing whether these projections are consistent with a single underlying $R_{\mu\nu}$ requires a subsequent, control-dependent inversion procedure, described below.

## 6.2 Model hierarchy and scope of the present work

The effective generators extracted in §6.1 can be interpreted within a hierarchy of models of increasing generality, reflecting different assumptions about the structure and time dependence of the relaxation tensor.

**Model 0 (baseline scalar models):** Conventional approaches treat relaxation using decoupled scalar rates, such as $R_1$ and $R_2$, possibly augmented by stretched exponentials or multi-exponential decompositions. These models assume that longitudinal and transverse dynamics factorize and that relaxation can be described independently of the control protocol. As discussed in §§4–5, this approximation breaks down in the presence of detuning, anisotropy, or noncommuting pulse sequences, where ordering sensitivity and basis motion couple relaxation channels.

**Model 1 (effective constant tensor; this work):** In the present work, we adopt an effective, cycle-averaged description in which the transverse dynamics are governed by a constant generator $A_{eff}$, and the dissipative component is represented by a symmetric tensor $R_{\perp}$. This model treats the measured relaxation as a sequence-dependent effective projection of an underlying covariant object, incorporating both intrinsic dissipation and control-induced mixing. The results shown in Fig. 4 correspond to this level of description, with separate effective tensors extracted for each pulse family and control setting.

**Model 2 (periodic or pulse index-dependent tensors):** A more general description allows the relaxation tensor to vary over the pulse cycle, e.g., $R(n)$ or piecewise-constant tensors defined during pulses and free-precession intervals. In this case, the observed dynamics arise from a periodic Liouvillian, and the effective generator corresponds to a Floquet (cycle-averaged) object obtained from the ordered product of segment-wise evolution operators. This model explicitly captures intra-cycle variations and higher-order commutator effects.

**Model 3 (covariant tensor with control-dependent projection):** At the most general level, relaxation is described by a covariant tensor $R_{\mu\nu}$ defined independently of the control sequence. The experimentally observed tensors are then obtained by applying a control-dependent map that accounts for basis transformations, time ordering, and geometric (holonomy) effects induced by the pulse sequence. In this framework, different pulse trains probe different projections of the same underlying object.

The present work operates at the level of Model 1, demonstrating that noncommuting control sequences produce distinct, reproducible effective projections that cannot be captured by scalar models. The broader objective is to test whether these projections are consistent with a single underlying covariant tensor, which would require constructing and inverting the corresponding control-dependent maps as outlined in §6.6.

### 6.3 Positive semi-definite formulation and physical constraints

To parameterize the dissipative component of the dynamics, we adopt a representation consistent with the Gorini–Kossakowski–Lindblad–Sudarshan (GKLS) form of the Liouvillian. In this framework, the dissipator can be expressed in an operator basis $\{F_\alpha\}$ as a quadratic form,

$$\mathcal{D}(\rho) = -\frac{1}{2}\sum_{\alpha,\beta} R_{\alpha\beta}\,[F_\alpha,[F_\beta,\rho]], \tag{S195}$$

where $R_{\alpha\beta}$ is a real, symmetric, positive semi-definite (PSD) matrix. The PSD property ensures complete positivity and corresponds physically to entropy production and contraction of distinguishability under open-system evolution.

When projected to the Bloch representation, the resulting generator acting on the magnetization vector generally contains three components: (i) a symmetric part, corresponding to dissipative contraction, (ii) an antisymmetric part, corresponding to coherent rotation, and (iii) an affine term, corresponding to drift toward equilibrium. Accordingly, the full $3 \times 3$ Bloch generator need not itself be symmetric or PSD. The PSD constraint applies specifically to the underlying dissipative quadratic form, not to the entire reduced generator in a particular representation.

In the present work, we extract the symmetric part of the effective transverse generator,

$$R_\perp = -\frac{1}{2}(A_{eff} + A_{eff}^T), \tag{S196}$$

and interpret it as an effective dissipative tensor within the observed $(S_x, S_y)$ subspace. This object inherits the role of the PSD dissipative sector, but only after projection and coarse-graining under the pulse sequence. As such, $R_\perp$ should be viewed as a sequence-dependent effective representation of the underlying PSD dissipative form, rather than a direct measurement of the microscopic Kossakowski matrix.

### 6.4 Fitting implementation

The fitting procedure proceeds by constructing effective transverse generators from the measured stroboscopic trajectories for each pulse train and control setting. For each dataset (defined by pulse family, transmitter offset, and pulse width), the transverse magnetization components $S_x(n)$ and $S_y(n)$ are first reconstructed as functions of pulse index $n$, which may be downsampled from millions of points to ~1000 points. These trajectories represent the state of the system immediately following each pulse, after demodulation of the free precession between pulses into a single complex point per cycle.

An effective linear generator $A_{eff}$ is then obtained by fitting the evolution of the transverse state across successive pulse indices according to

$$\frac{d}{dn}\begin{pmatrix} S_x \\ S_y \end{pmatrix} = A_{eff}\begin{pmatrix} S_x \\ S_y \end{pmatrix}. \tag{S197}$$

In practice, this corresponds to identifying the constant $2 \times 2$ matrix that best describes the local linear mapping between successive points along the trajectory. This procedure yields a cycle-averaged generator that incorporates both coherent evolution (including detuning) and dissipative effects under the applied pulse sequence.

The dissipative component is then extracted as the symmetric part,

$$R_{\perp} = -\frac{1}{2}(A_{eff} + A_{eff}^{T}), \tag{S198}$$

and the tensor elements $R_{11}, R_{12}, R_{21}, R_{22}$ are recorded for each dataset. These values define the effective transverse relaxation tensor for the given control setting.

For each pulse family, the fitting is performed independently across systematic sweeps of transmitter offset and pulse width. The results shown in Fig. 4 of the main text are obtained by averaging the fitted tensor components over pulse width at fixed offset, providing a reduced representation of the dependence on detuning.

Three pulse-train families are considered: a commuting reference train (zero winding) and two noncommuting phase-wound trains (positive and negative winding). The commuting train samples a nearly fixed axis in Liouville space and yields an effective tensor that is close to diagonal in the detection basis. In contrast, the phase-wound trains repeatedly cycle through noncommuting control directions, producing effective generators that incorporate ordering-dependent contributions. The resulting tensors therefore exhibit off-diagonal structure and systematic dependence on both the magnitude and sign of the transmitter offset.

The fitted tensors obtained in this manner are interpreted as effective Floquet (cycle-averaged) projections of the underlying dissipative dynamics. In the commuting case, this projection approximates a simple dressed-to-laboratory frame transformation. In the noncommuting case, however, the effective generator arises from the full time-ordered product of coherent and dissipative evolution over the pulse cycle, and therefore mixes intrinsic relaxation, control-induced frame transport, and geometric (holonomy) effects.

## 6.5 Failure modes and limitations

Fitting effective generators from stroboscopic trajectories is an exercise subject to both experimental and structural failure modes. These arise primarily from ambiguity between coherent phase evolution and dissipative decay, as well as from limitations in signal quality and sampling. A common experimental issue is the appearance of phase instabilities or discretization artifacts (e.g., phase "bin-hopping"), in which small errors in phase unwrapping or demodulation produce discontinuous jumps in the reconstructed trajectory. These artifacts can lead to spurious oscillations or apparent high-frequency structure in the fitted generator if not properly identified and excluded. Because the present approach relies on local linear structure in the $(S_x, S_y)$ trajectories, such discontinuities must be removed or masked prior to fitting.

More fundamentally, identifiability requires a clear separation between coherent and dissipative contributions to the evolution. The transverse dynamics include both detuning-driven phase accumulation and relaxation-induced contraction. If coherent phase evolution is not explicitly accounted for, it can be incorrectly absorbed into the symmetric part of the fitted generator, leading to artificial off-diagonal structure or incorrect decay rates. For this reason, the generator is decomposed into symmetric (dissipative) and antisymmetric (coherent) components, and only the symmetric part is interpreted as the relaxation tensor.

Another limitation can arise from insufficient sampling of independent directions in Liouville space. Commuting pulse trains probe a restricted set of directions and therefore provide limited sensitivity to off-diagonal tensor components. In such cases, the fitted generator may appear nearly diagonal even if the underlying relaxation tensor contains anisotropy. Noncommuting (phase-wound) pulse trains mitigate this limitation by enforcing exploration of multiple noncollinear directions, thereby improving identifiability through ordering sensitivity.

Finally, the effective generator extracted here is a cycle-averaged quantity. When intra-cycle dynamics vary strongly (e.g., between pulse and free precession intervals), a constant generator provides only an approximation. This limitation is addressed by extending to periodic or segment-resolved models (Model 2), but is not required to capture the dominant features observed in the present data.

These considerations emphasize that relaxation tensor fitting is an operator identification problem: meaningful extraction requires separating coherent and dissipative structure, ensuring sufficient directional sampling through control design, and maintaining data quality in the reconstructed trajectories.

## 6.6 Global covariant reconstruction and Floquet inversion

The effective tensors extracted in §6.4 represent sequence-dependent projections of an underlying dissipative structure. A central question is whether these projections are consistent with a single covariant relaxation tensor $R_{\mu\nu}$, independent of the control sequence, and if so, how that tensor may be reconstructed. To address this, we distinguish two stages of analysis.

**Stage 1: empirical, family-specific fits:** In the present work, each pulse-train family (commuting zero-winding and noncommuting positive and negative winding) is fit independently across systematic sweeps of transmitter offset and pulse width. These fits yield effective transverse tensors $R_\perp$ that vary with control parameters. The primary observation is that commuting and noncommuting sequences produce distinct but reproducible tensor structures: commuting trains yield near-diagonal projections, while noncommuting trains reveal off-diagonal components and asymmetries that depend on control history. This stage establishes that the measured dynamics cannot be described by a single scalar relaxation rate and are consistent with a tensorial description.

**Stage 2: global covariant reconstruction:** A stronger test of the tensor framework is to determine whether all measured projections can be explained by a single underlying covariant tensor $R_{\mu\nu}$. This requires constructing a sequence-specific forward model that maps the covariant dissipative form into the experimentally observed effective generators.

For a given pulse sequence $p$, the effective generator may be written schematically as

$$A_{eff}^{(p)} = \mathcal{F}_p \left[ \mathcal{L}_H(t), \mathcal{D}(R_{\mu\nu}) \right], \tag{S199}$$

where $\mathcal{L}_H(t)$ is the control-dependent Hamiltonian Liouvillian and $\mathcal{D}(R_{\mu\nu})$ is the dissipative superoperator generated by the covariant tensor. The map $\mathcal{F}_p$ represents the time-ordered (Floquet) composition of coherent and dissipative evolution over one pulse cycle, including higher-order commutator (Magnus) contributions when the generators do not commute. As developed with Killing vector formalism in §3, the pulse train is a gauge choice in Liouville space, the control contributes a connection, and the effective generator over a cycle is obtained by a time-ordered exponential in the Magnus expansion, with the second-order commutator term representing curvature and holonomy.

In the commuting limit, where the control effectively samples a single direction in Liouville space, $\mathcal{F}_p$ reduces approximately to a basis transformation (e.g., dressed-to-laboratory frame rotation), and the measured tensor is close to a simple projection of $R_{\mu\nu}$. In contrast, for noncommuting pulse trains, $\mathcal{F}_p$ is a genuinely nontrivial map that mixes dissipation, frame transport, and geometric (holonomy) effects through time ordering. In this regime, the effective tensor is a cycle-averaged (Floquet) object rather than a direct instantaneous representation.

The covariance hypothesis can then be formulated as the existence of a single $R_{\mu\nu}$ such that

$$R_{\perp}^{(p)} \;\approx\; \mathcal{P}_{\perp}\left(\mathcal{F}_p[R_{\mu\nu}]\right) \tag{S200}$$

for all pulse families $p$, where $\mathcal{P}_{\perp}$ denotes projection onto the observed transverse subspace. Testing this hypothesis requires:

1. Specifying or simulating the control-dependent maps $\mathcal{F}_p$ for each pulse sequence,
2. Identifying a candidate covariant tensor $R_{\mu\nu}$, and
3. Comparing the predicted effective projections with the independently fitted tensors from Stage 1.

If a consistent $R_{\mu\nu}$ can be found, the result would demonstrate that the sequence-dependent effective tensors are unified by a single underlying dissipative object. Conversely, failure of such a reconstruction would indicate that additional structure (e.g., time dependence, higher-order coupling, or model mismatch) must be incorporated. The present work establishes the first stage of this program by demonstrating systematic, control-dependent tensor projections. Implementation of the full inversion, including explicit construction of $\mathcal{F}_p$ for each pulse family, constitutes a natural extension of this framework.

## 6.7 NNLS modal decomposition (secondary analysis)

As an independent comparison, relaxation dynamics were also analyzed using a nonnegative least squares (NNLS) decomposition of the signal into a basis of exponential modes. In this approach, the transverse signal is represented as

$$S(n) = \sum_k w_k \, e^{-n/\tau_k}, \tag{S201}$$

where $\{\tau_k\}$ span a predefined range of decay times and the weights $w_k \geq 0$ are determined by NNLS inversion. This provides a model-free estimate of the distribution of relaxation timescales contributing to the observed signal.

The NNLS decomposition is useful for visualizing multicomponent relaxation behavior and for identifying distinct dynamical regimes across control parameters such as transmitter offset and pulse width. In particular, the resulting maps (e.g., Fig. S2) reveal structured variation in dominant timescales and mode weights that correlate with the control sequence, indicating that the relaxation dynamics cannot be described by a single characteristic rate.

However, this approach does not directly preserve operator-level structure. The NNLS representation decomposes the signal into scalar decay components and therefore does not retain information about directional coupling between $S_x$ and $S_y$, nor does it distinguish between coherent and dissipative contributions to the evolution. As a result, it cannot be used to extract the effective generator $A_{eff}$ or the transverse tensor $R_\perp$. For this reason, NNLS analysis is used here as a diagnostic and complementary tool, rather than as the primary method for tensor extraction. The effective transverse tensors reported in the main text are obtained from direct fitting of the transverse trajectories (§6.4), which preserves the operator structure necessary to resolve anisotropy and off-diagonal coupling.

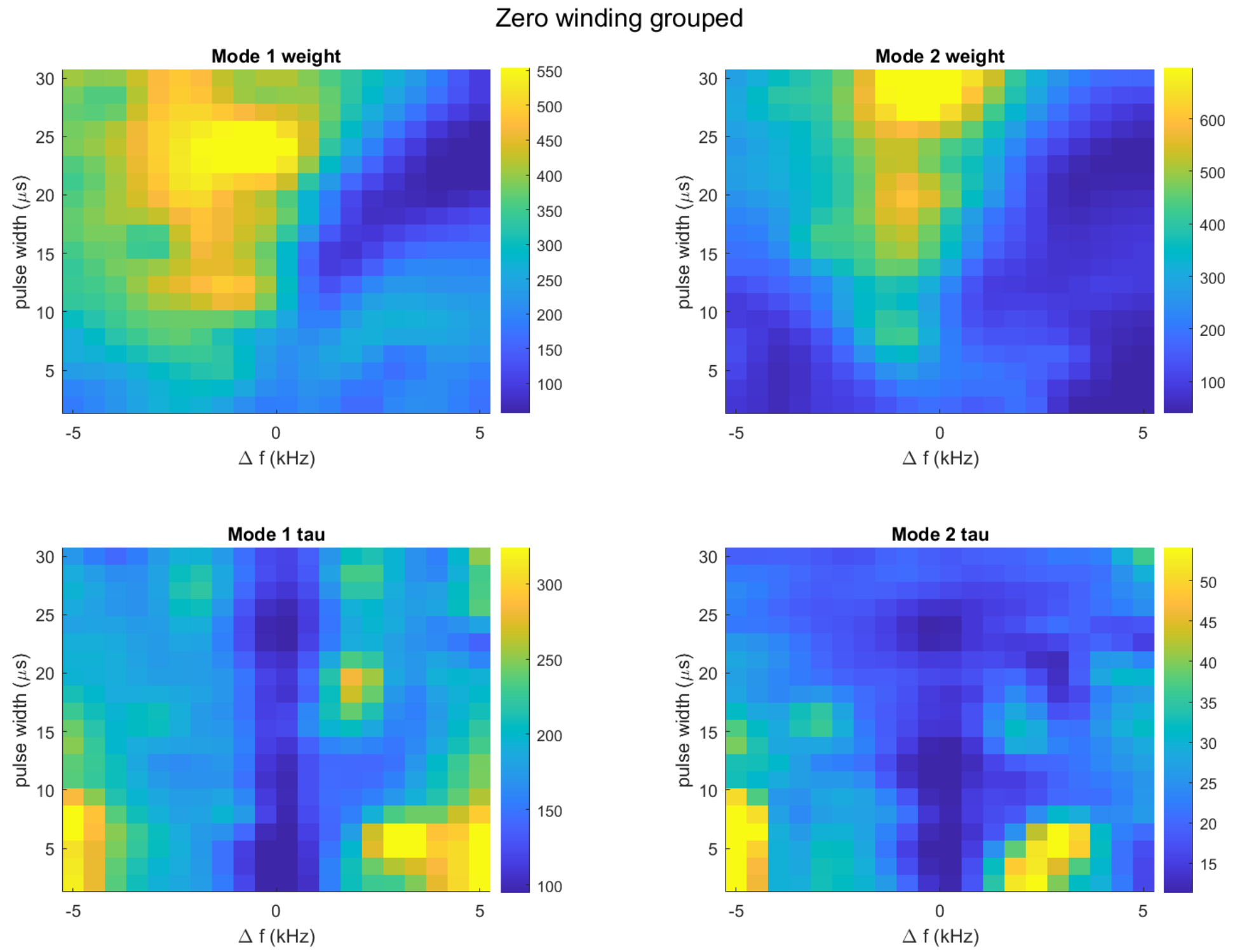

Fig. S2: **NNLS modal decomposition of commuting pulse trains.** NNLS modal decomposition for the commuting (zero-winding) pulse trains, shown here for $^{13}$C nuclear spins in diamond with NV centers. The mode weights are shown in **(a)** and **(b)**, respectively, and tau (inverse rates) are shown in **(c)** and **(d)**, respectively in seconds. To broadly examine how the effective modes vary across control parameters using a model-free approach, we map the fitted components over transmitter frequency offset and pulse width. These measurements sample a two-dimensional control space in which both the effective rotation axis and the relative weighting of noncommuting contributions are varied. Here, NNLS inversion was performed using basis sets of 80 exponentials with rates spanning three orders of magnitude, then decomposed into the dominant modes.

## Supplementary References